\documentclass[reprint,prl,aps,longbibliography]{revtex4-2}
\usepackage{times}
\usepackage{graphicx}
\usepackage{amsmath, braket, amsfonts}
\usepackage{amssymb}
\usepackage{sidecap}
\usepackage{bm, color, ulem}
\usepackage{xcolor}
\usepackage{ragged2e}
\usepackage{float}
\usepackage{wrapfig,lipsum,booktabs}
\usepackage[colorlinks=true, allcolors=blue]{hyperref}
\usepackage{cleveref}
\usepackage{siunitx}
\renewcommand{\figurename}{\textbf{Fig.}} 

\newcommand{\be}{\begin{equation}}
\newcommand{\ee}{\end{equation}}

\DeclareUnicodeCharacter{03BD}{{$\nu$}}
\DeclareUnicodeCharacter{2212}{{-}}

\makeatletter
\def\maketitle{
\@author@finish
\title@column\titleblock@produce
\suppressfloats[t]}
\makeatother

\begin{document}
\title{Strain-tunable anomalous Hall effect in hexagonal MnTe}

\author{Zhaoyu Liu$^{1,2}$}
\thanks{Authors contributed equally to this work.}
\email{\\liuzy.phy@gmail.com}
\author{Sijie Xu$^{1,2}$} 
\thanks{Authors contributed equally to this work.}
\author{Jonathan M. DeStefano$^{3}$} 
\author{Elliott Rosenberg$^{3}$} 
\author{Tingjun Zhang$^{1,2}$} 
\author{Jinyulin Li$^{4}$} 
\author{Matthew B. Stone$^{5}$} 
\author{Feng Ye$^{5}$} 
\author{Wei Tian$^{5}$} 
\author{Sarah Edwards$^{3}$}
\author{Rong Cong$^{6}$} 
\author{Siyu Pan$^{1,2}$} 
\author{Ching-Wu Chu$^4$} 
\author{Liangzi Deng$^{4}$} 
\author{Emilia Morosan$^{1,2}$} %
\author{Rafael M. Fernandes$^{7,8}$}
\author{Jiun-Haw Chu$^{3}$}
\email{jhchu@uw.edu}
\author{Pengcheng Dai$^{1,2}$}
\email{pdai@rice.edu}

\affiliation{$^{1}$Department of Physics \& Astronomy, Rice University, Houston, TX 77005, USA}
\affiliation{$^{2}$Rice Laboratory for Emergent Magnetic Materials and Smalley-Curl Institute, Rice University, Houston, TX 77005, USA}
\affiliation{$^{3}$Department of Physics, University of Washington, Seattle, WA 98195, USA}
\affiliation{$^{4}$Department of Physics and Texas Center for Superconductivity, University of Houston, Houston, TX 77204, USA}
\affiliation{$^{5}$Neutron Scattering Division, Oak Ridge National Laboratory, Oak Ridge, TN, USA}
\affiliation{$^{6}$National High Magnetic Field Laboratory, Florida State University, Tallahassee, Florida 32306, USA}
\affiliation{$^{7}$Department of Physics, The Grainger College of Engineering, University of Illinois Urbana-Champaign, Urbana, Illinois 61801, USA}
\affiliation{$^{8}$Anthony J. Leggett Institute for Condensed Matter Theory, The Grainger College of Engineering, University of Illinois Urbana-Champaign, Urbana, Illinois 61801, USA}


\begin{abstract}
The ability to control and manipulate time-reversal ($T$) symmetry-breaking phases with near-zero net magnetization is a sought-after goal in spintronic devices  \cite{RevModPhys.76.323,RevModPhys.90.015005}. The recently discovered hexagonal altermagnet manganese telluride ($\alpha$-MnTe) is a prime example \cite{doi:10.1126/sciadv.aaz8809,PhysRevX.12.031042,PhysRevX.12.040501,WOS:000775805300001,daiResearchProgressFuture2025,fenderAltermagnetismChemicalPerspective2025,WOS:001147635900001}. It has a compensated altermagnetic ground state where the magnetic moments are aligned in each layer and stacked antiparallel along the $c$-axis \cite{kunitomiNeutronDiffractionStudy1964,efremdsaLowtemperatureNeutronDiffraction2005,szuszkiewiczNeutronScatteringStudy2005,kriegnerMagneticAnisotropyAntiferromagnetic2017,szuszkiewiczSpinwaveMeasurementsHexagonal2006}, yet it exhibits a spontaneous anomalous Hall effect (AHE) that breaks the $T$-symmetry with a vanishingly small $c$-axis ferromagnetic (FM) moment \cite{krempaskyAltermagneticLiftingKramers2024,aminNanoscaleImagingControl2024,PhysRevLett.133.156702,gonzalezbetancourtSpontaneousAnomalousHall2023,kluczykCoexistenceAnomalousHall2024,leeBrokenKramersDegeneracy2024,loveseyTemplatesMagneticSymmetry2023,mazinAltermagnetismMnTeOrigin2023,PhysRevLett.132.236701,PhysRevLett.133.086503,839n-rckn}. However, the presence of three 120$^\circ$ separated in-plane magnetic domains \cite{kunitomiNeutronDiffractionStudy1964,efremdsaLowtemperatureNeutronDiffraction2005,szuszkiewiczNeutronScatteringStudy2005,kriegnerMagneticAnisotropyAntiferromagnetic2017,szuszkiewiczSpinwaveMeasurementsHexagonal2006} presents a challenge in understanding the origin of the AHE and the effective control of the altermagnetic state \cite{gonzalezbetancourtSpontaneousAnomalousHall2023,kluczykCoexistenceAnomalousHall2024}. Here we use neutron scattering to show that symmetry breaking anisotropic strain, induced by compressive uniaxial pressure along the nearest-neighbor (NN) Mn--Mn bond directions, detwins $\alpha$-MnTe into a single in-plane magnetic domain. This control over in-plane domains allows us to unambiguously establish that the in-plane moments are aligned along the NNN Mn--Mn bond direction, irrespective of the applied strain directions \cite{gonzalezbetancourtSpontaneousAnomalousHall2023,kluczykCoexistenceAnomalousHall2024}. Mounting the sample on a piezoelectric strain cell along both NN and NNN directions can drive the sample into a single-domain state that significantly sharpens the AHE hysteresis loop and extends the AHE to lower temperatures. Furthermore, tuning the uniaxial strain reverses the sign of the AHE near room temperature. Remarkably, this is achieved without altering the altermagnetic phase-transition temperature or substantially changing the small $c$-axis FM moment. Combined with our phenomenological model, we argue that these effects result from the modification of the electronic Berry curvature by a combination of both spin-orbit coupling and strain  \cite{belashchenkoGiantStrainInducedSpin2025,WOS:001530763600001}. Our work not only unambiguously establishes the relationship between the in-plane moment direction and the AHE in $\alpha$-MnTe but also paves the way for future applications in highly scalable, strain-tunable magnetic sensors and spintronic devices \cite{RevModPhys.76.323,RevModPhys.90.015005,PhysRevX.12.040501,WOS:000775805300001,daiResearchProgressFuture2025,fenderAltermagnetismChemicalPerspective2025,WOS:001147635900001,bangarInterplayAltermagneticOrder2025,PhysRevB.108.075425,Jungwirth_Fernandes2025,Jungwirth2025symmetry}.
\end{abstract}
\maketitle
\begin{figure*}[ht!]
    \includegraphics[width = 180mm]{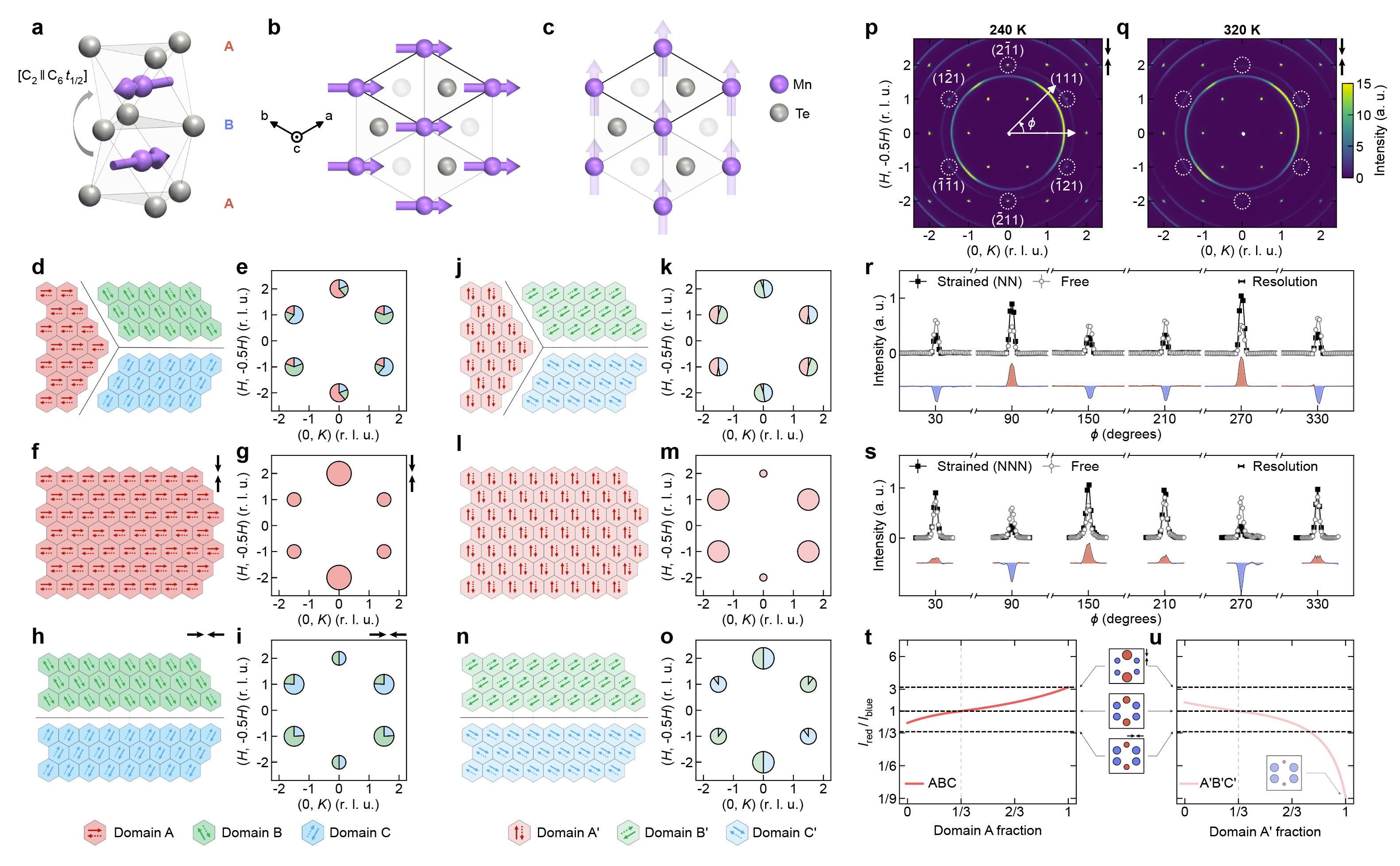}
    \caption{
    \textbf{Neutron scattering experiments under uniaxial strain.}
    \textbf{a,} Crystal structure of $\alpha$-MnTe.
    \textbf{b,} In-plane magnetic structure with the Mn$^{2+}$ moments along the next-nearest-neighbor (NNN) Mn--Mn bond direction, corresponding to $\left[1 \bar{1} 0 \right]$.
    \textbf{c,} Similar to \textbf{b} with the Mn$^{2+}$ moment along the nearest-neighbor (NN) Mn--Mn bond direction, corresponding to $\left[1 1 0 \right]$.
    \textbf{d-i} Schematics of the magnetic domains and corresponding simulated Bragg-intensity pie charts for Mn moments along the NNN Mn--Mn bond direction (as in \textbf{b}, yielding three in-plane magnetic domains labeled as A, B, C). (\textbf{d}) In the free-standing crystal, six symmetry-equivalent domains are uniformly populated, each color represents a pair of time-reversal-related domains $\rightleftarrows$ and $\leftrightarrows$ with the same in-plane moment direction, which in turn can be along one of three directions related by threefold rotation. (\textbf{f}) Under compressive uniaxial strain along the NN Mn--Mn bond direction the sample is detwinned, leaving a single pair of domains with moments aligning perpendicular to the uniaxial strain direction. 
    (\textbf{h}) Under compressive uniaxial strain along the NNN Mn--Mn bond direction, the sample becomes partially detwinned: the domain with moments parallel to the strain axis is fully suppressed, leaving only the two remaining in-plane domain orientations.
    \textbf{j-o,} Schematics and simulations analogous to \textbf{d–i}, but for Mn$^{2+}$ moments aligned along the NN Mn--Mn bond direction (as shown in \textbf{c}, yielding three in-plane magnetic domains labeled as A$^\prime$, B$^\prime$, and C$^\prime$). Panels \textbf{j}, \textbf{l}, and \textbf{n} present symmetry-allowed domain configurations (under strain) that qualitatively reproduce the intensity patterns. Details of the domain selection inferred from the strained data are provided in Methods.
    \textbf{p,q,} Elastic neutron scattering intensity maps in the $HK$ plane at $L=1$ (r.l.u.) under compressive uniaxial strain  along the NN Mn--Mn bond direction at 240 K and 320 K. Dashed white circles mark the magnetic peaks positions. The white arrow defines the azimuthal angle $\phi$ in the following panel.
    \textbf{r,} Azimuthal angular dependence of magnetic peak intensities for the free-standing and NN-strained samples, along with their difference. Color shading highlights that the two peaks perpendicular to the uniaxial-stress direction are enhanced (red), while the other four are suppressed (blue).
    \textbf{s,} Azimuthal angular dependence of magnetic peak intensities for the free-standing and NNN-strained samples, measured in a separate neutron experiment with uniaxial stress applied along the NNN Mn--Mn bond direction. All panels \textbf{p-s} are at $L=1$ (r.l.u.).
    \textbf{t,} Calculated diffraction-intensity ratio $I_{\mathrm{red}}/I_{\mathrm{blue}}$ versus the volume fraction of Domain A within the $m^\prime m^\prime m$ magnetic point group. The ratio is defined as the averaged integrated intensity of one symmetry-related set of magnetic peaks (red) divided by that of the complementary set (blue). The black dash lines mark the experimentally determined ratio. 
    \textbf{u,} Corresponding calculation for Domain A$^\prime$, B$^\prime$, and C$^\prime$ within the $mmm$ magnetic point group.
    }
    \label{fig:fig1}
\end{figure*}

Altermagnets are collinear magnetic states with compensated moments and ‘alter’nating orientations of local crystalline environments, possessing a joint time-reversal ($T$) and crystalline rotational symmetry that enables ferromagnetic (FM) like behaviors (nonzero spin-splitting) but with vanishingly small net magnetization \cite{doi:10.1126/sciadv.aaz8809,PhysRevX.12.031042,PhysRevX.12.040501,WOS:000775805300001,daiResearchProgressFuture2025,fenderAltermagnetismChemicalPerspective2025,WOS:001147635900001,Jungwirth_Fernandes2025,Jungwirth2025symmetry}. The rotational symmetry that relates the sublattices hosting the opposite magnetic moments in altermagnets gives rise to spin-splitting in the energy bands with even-parity $d$-, $g$-, or
$i$-wave symmetry even without relativistic spin-orbit coupling (SOC). With finite SOC, as is generally the case for real materials, a nonzero net magnetic moment and a finite anomalous Hall effect (AHE) can emerge for certain moment orientations \cite{doi:10.1126/sciadv.aaz8809,PhysRevX.12.031042,PhysRevX.12.040501,antonenkoMirrorChernBands2025,WOS:000775805300001,daiResearchProgressFuture2025,fenderAltermagnetismChemicalPerspective2025,WOS:001147635900001}. For altermagnets such as hexagonal manganese telluride ($\alpha$-MnTe), whose antiparallel spins are connected by a screw symmetry operation \cite{krempaskyAltermagneticLiftingKramers2024,aminNanoscaleImagingControl2024,PhysRevLett.133.156702,gonzalezbetancourtSpontaneousAnomalousHall2023}, the momentum ($\mathbf{k}$)-resolved Berry curvature \cite{PhysRevLett.88.207208,RevModPhys.82.1539} depends sensitively on the direction of the in-plane moments relative to the underlying crystalline lattice (Figs. \ref{fig:fig1}a-c) \cite{gonzalezbetancourtSpontaneousAnomalousHall2023,kluczykCoexistenceAnomalousHall2024}. If the in-plane moments are aligned along the next-nearest-neighbor (NNN) Mn--Mn bond direction (Fig.~\ref{fig:fig1}b) a finite AHE is symmetry allowed, which could be generated from the nonzero integration of the $\mathbf{k}$-dependent Berry curvature. In contrast, no AHE is expected if the moments are aligned along the nearest-neighbor (NN) Mn--Mn bond  (Fig.~\ref{fig:fig1}c) \cite{gonzalezbetancourtSpontaneousAnomalousHall2023,kluczykCoexistenceAnomalousHall2024}. However, in free-standing $\alpha$-MnTe, neutron diffraction experiments cannot conclusively determine the direction of the in-plane moments owing to three 120$^\circ$ separated in-plane magnetic domains. Domain averaging yields identical in-plane magnetic scattering intensity distributions for moments oriented along either the NNN or NN Mn--Mn bond direction (Figs. \ref{fig:fig1}d,e,j,k)\cite{kunitomiNeutronDiffractionStudy1964,efremdsaLowtemperatureNeutronDiffraction2005,szuszkiewiczNeutronScatteringStudy2005,kriegnerMagneticAnisotropyAntiferromagnetic2017,szuszkiewiczSpinwaveMeasurementsHexagonal2006}. 

By carrying out neutron-scattering experiments under uniaxial compressive strain applied either along the NN or NNN Mn–Mn bond direction, we observe two complementary patterns of Bragg-intensity redistribution in $\alpha$-MnTe. The opposite strain responses are only compatible with the detwinning process expected for a magnetic configuration in which the Mn moments lie along the NNN Mn–Mn bond direction, and this easy axis is unaffected by strain directions (Figs. \ref{fig:fig1}t,u and Extended Data Figs. 2-4).  These results therefore directly establish the link between the in-plane moment orientation and the AHE \cite{gonzalezbetancourtSpontaneousAnomalousHall2023,kluczykCoexistenceAnomalousHall2024}.

\begin{figure}[!b]
    \includegraphics[width = 86mm]{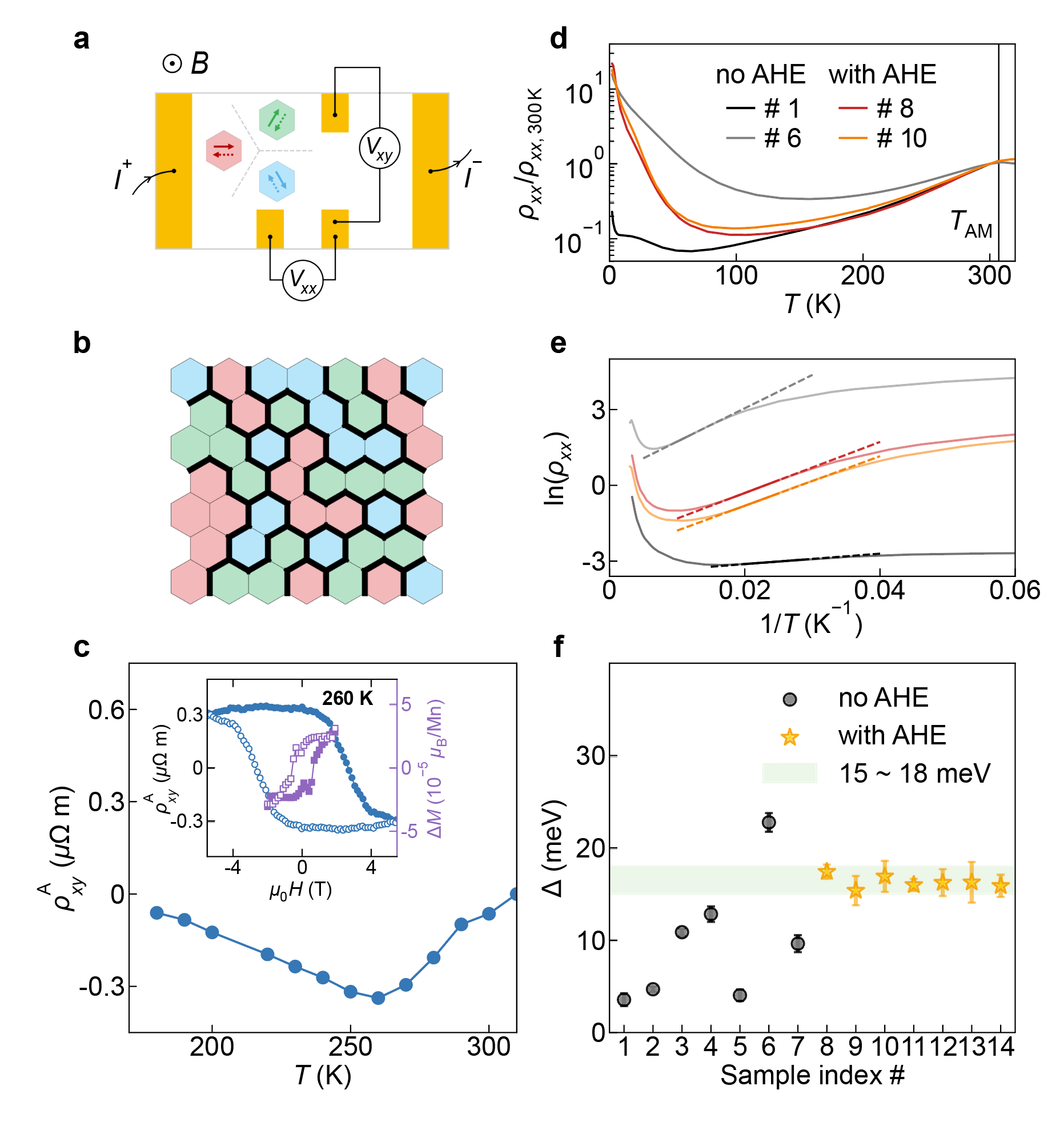}
    \caption{
    \textbf{The AHE measurements in free-standing $\alpha$-MnTe.}
    \textbf{a,} Schematic of the five-terminal Hall bar used for electrical transport measurements. The current is applied along either the NN or NNN Mn--Mn bond direction, and the magnetic field is applied along the $c$ axis. Colored hexagons indicate the in-plane magnetic-domain orientations.
    \textbf{b,} Illustration of uniformly distributed magnetic domains and domain boundaries.
    \textbf{c,} $\rho_{xy}^{\mathrm A}$ as function of temperature. Inset: Hall resistivity and magnetization hysteresis loops at 260 K after linear background subtraction.
    \textbf{d,} Temperature dependence of the normalized longitudinal resistivity $\rho_{xx}$ for samples grown by different methods. Samples \#8 and \#10, grown by flux method, exhibit AHE, whereas Samples \#1 and \#6, grown by chemical vapor transport and flux methods with different procedures, show no AHE. The vertical line marks the altermagnetic transition temperature ($T_{\mathrm{AM}}$). \textbf{e}, Arrhenius plot for extracting the thermal activation energy, with the same data in \textbf{d}. The dashed lines represent linear fits based on $\rho_{xx} \propto e^{\Delta/2k_{\mathrm B}T}$, where $k_{\mathrm B}$~is the Boltzmann constant and $\Delta$ is the activation energy gap, i.e. the charge gap.
    \textbf{f,} Summary of the charge gap for various samples. All samples exhibiting AHE (yellow stars) lie within a narrow range of 15--18 meV. 
    \label{fig:fig2}}
\end{figure}

\begin{figure*}[ht!]
    \includegraphics[width = 180mm]{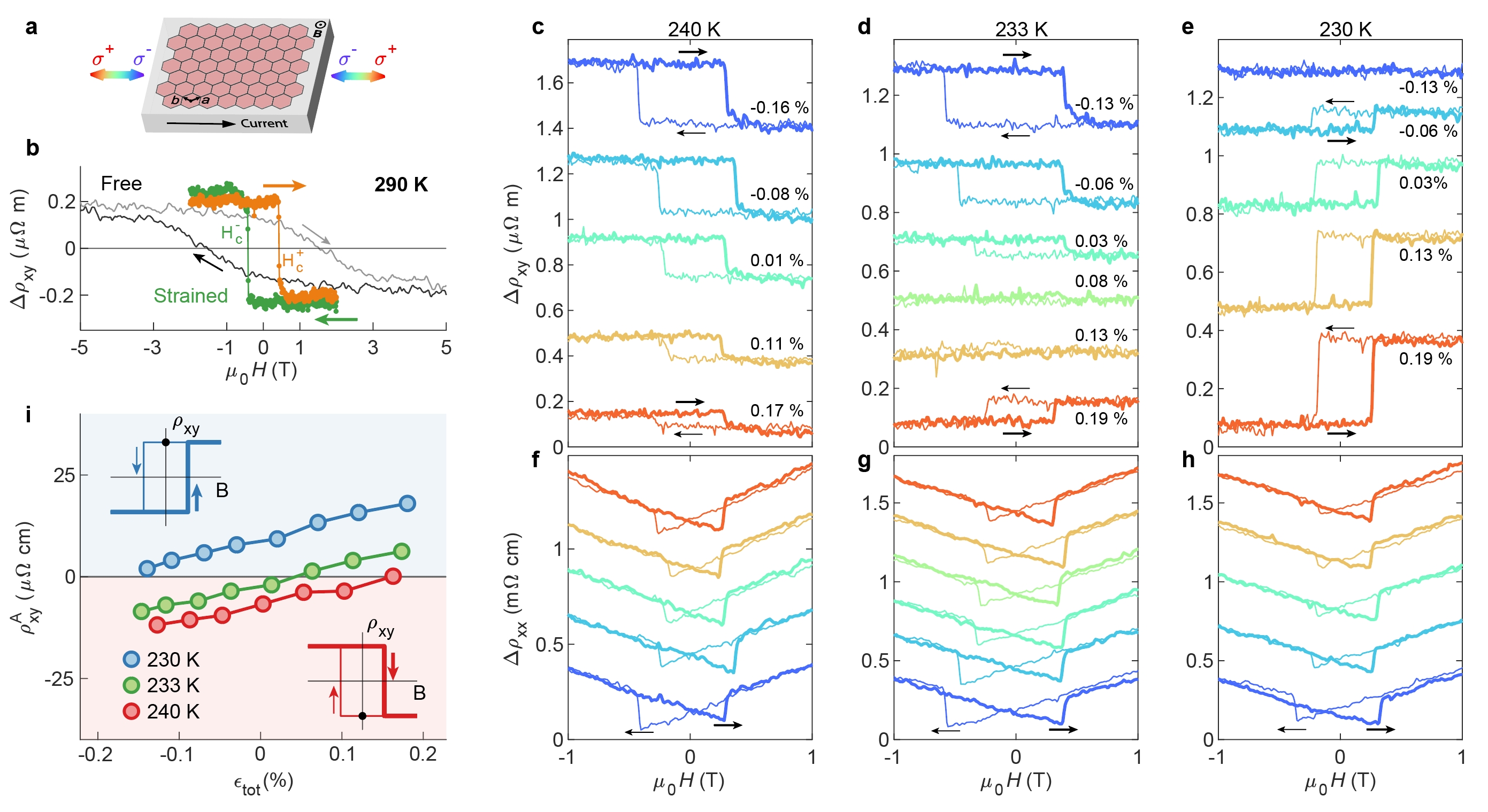}
    \caption{
    \textbf{Sign reversal of the AHE in $\alpha$-MnTe driven by uniaxial strain and temperature.}
    \textbf{a,} Schematic showing the orientation of uniaxial stress applied along the crystal NNN Mn--Mn bond direction in strained transport measurements. Blue arrows represent compressive stress $\sigma^{-}$, while red arrows indicate tensile stress $\sigma^{+}$. The electrical current is along the same direction. Magnetic field is applied out-of-plane.
    \textbf{b,} Comparison of Hall resistivity $\Delta\rho_{xy}$ (linear background subtracted) as a function of magnetic field between a sample in the free-standing state and strained state at 290 K. Arrows denote the magnetic-field sweep directions. $H_c^{+}$ and $H_c^{-}$ represent the coercive fields.
    \textbf{c-e,} $\Delta\rho_{xy}$ (linear background subtracted) at 240, 233 and 230~K under different strain levels; the legend denotes the total strain experienced by the crystal (including both the mechanically applied strain and the thermally induced strain). Thin (thick) black arrows indicate field sweeps down (up). Curves are offset for clarity.
    \textbf{f-h,} Corresponding longitudinal resistivity $\rho_{xx}$ measured under the same strain and temperature conditions as in \textbf{b-d}. The data have been shifted vertically by subtracting constant offsets.
    \textbf{i,} Extracted anomalous Hall resistivity $\rho_{xy}^{\mathrm A}$ versus total strain at different temperatures. The red shading denotes negative $\rho_{xy}^{\mathrm A}$, corresponding to the black dot in the bottom-right inset, whereas the light-blue shading indicates positive $\rho_{xy}^{\mathrm A}$, corresponding to the black dot in the top-left inset.}
    \label{fig:fig3}
\end{figure*}

Since $\alpha$-MnTe is a self-doped semiconductor with residual resistivity sensitively dependent on sample preparation methods, we carried out systematic transport measurements on free-standing single crystals with different residual resistivity and find that the AHE only appears for samples with a charge gap of $\sim$ 15--18 meV (Fig.~\ref{fig:fig2}). Furthermore, we find that anisotropic strain ($\epsilon$) controls the sign of the AHE near room temperature (Fig.~\ref{fig:fig3} and Extended Data Figs. 6-8). 
Since strain does not change in the in-plane moment direction relative to the underlying crystalline lattice  \cite{Coey_2010} nor the switching of the magnetic moment along the $c$ axis \cite{gonzalezbetancourtSpontaneousAnomalousHall2023,kluczykCoexistenceAnomalousHall2024},  the observed sign change of the AHE (Fig.~\ref{fig:fig3}) cannot be explained by the piezomagnetic effect \cite{ikhlas2022piezomagnetic}. Instead, it is best attributed to strain-induced changes in electronic structure and $\mathbf{k}$-resolved Berry curvature~\cite{osin2025extrinsic}. Theoretical results from our phenomenological model are consistent with this interpretation, revealing that the combination of strain and SOC leads to an additional contribution to the AHE that is linear in strain and in the altermagnetic order parameter.  Our Hall resistivity measurements across varying temperatures and strain levels applied along the NNN and NN directions reveal that the AHE in strained $\alpha$-MnTe persists over a substantially broader temperature range and exhibits much sharper hysteresis loops than in free-standing crystal (Fig.~\ref{fig:fig4} and Extended Data Fig. 6). Because our elastocaloric (EC) effect measurements show no evidence of additional phase transitions beyond the altermagnetic one at $T_{\mathrm{AM}}$ near room temperature (Fig.~\ref{fig:fig4}o) \cite{ikedaACElastocaloricEffect2019,PhysRevX.14.031015}, the dramatic widening of the AHE temperature window in strained $\alpha$-MnTe is unlikely to originate from a new phase. Instead, we attribute it to the single-domain state stabilized by anisotropy strain, which enhances the net electronic Berry curvature and its coupling with magnetic fields by removing the twin-domain boundaries that do not contribute to the AHE (Fig.~\ref{fig:fig4}). Given that the strain-induced AHE and $T$-symmetry breaking in $\alpha$-MnTe occur near room temperature (Figs. \ref{fig:fig3} and \ref{fig:fig4}), $\alpha$-MnTe can be used as a strain-tunable magnetic sensor and spintronic device when integrated with other superconductors and topological insulators \cite{RevModPhys.76.323,RevModPhys.90.015005,PhysRevX.12.040501,WOS:000775805300001,daiResearchProgressFuture2025,fenderAltermagnetismChemicalPerspective2025,WOS:001147635900001,bangarInterplayAltermagneticOrder2025,PhysRevB.108.075425,Jungwirth_Fernandes2025,Jungwirth2025symmetry}

\section{Neutron Scattering under uniaxial compressive strain}

Figures \ref{fig:fig1}a-c show the compensated magnetic structure of $\alpha$-MnTe below $T_{\mathrm{AM}}\approx 307$ K with the space group $P6_3/mmc$ and the NiAs structure, in which Mn$^{2+}$ spins are parallel within the $ab$ plane and antiparallel along the $c$-axis (Figs. \ref{fig:fig1}a-c) \cite{kunitomiNeutronDiffractionStudy1964,efremdsaLowtemperatureNeutronDiffraction2005,szuszkiewiczNeutronScatteringStudy2005,kriegnerMagneticAnisotropyAntiferromagnetic2017,szuszkiewiczSpinwaveMeasurementsHexagonal2006}. 
The opposite spins are located in two atomic positions that are related by a screw symmetry (a sixfold rotation followed by a half translation), making this ordered state an altermagnet \cite{PhysRevX.12.031042}. 
In the NiAs structure, anions form a hexagonal close-packed lattice and cations occupy octahedral sites between the anion layers (Figs.~\ref{fig:fig1}a,b). By symmetry of the layers and the $ABAB$ stacking of the triangular anion lattice (Figs.~\ref{fig:fig1}a–c), free-standing $\alpha$-MnTe in the altermagnetic state has six symmetry-equivalent domains: three in-plane domains related by threefold rotation with moment orientations separated by $120^\circ$ (Figs.~\ref{fig:fig1}d or j), and, within each, two $c$-axis altermagnetic stacking variants $\rightleftarrows$ and $\leftrightarrows$ related by time reversal, as shown in Fig.~\ref{fig:fig1}a
\cite{leeBrokenKramersDegeneracy2024,loveseyTemplatesMagneticSymmetry2023,mazinAltermagnetismMnTeOrigin2023,PhysRevLett.132.236701,PhysRevLett.133.086503,839n-rckn,aoyamaPiezomag2024,beyUnexpectedTuningAnomalous2024}. Within the three 120$^\circ$ separated magnetic domains, SOC determines the in-plane easy axes relative to the underlying lattice \cite{Coey_2010}. However, domain averaging in free-standing $\alpha$-MnTe prevents prior neutron diffraction experiments from identifying whether the realized orientation is along the NNN (Figs.~\ref{fig:fig1}b) or the NN Mn--Mn bond (Figs.~\ref{fig:fig1}c) \cite{kunitomiNeutronDiffractionStudy1964,efremdsaLowtemperatureNeutronDiffraction2005,szuszkiewiczNeutronScatteringStudy2005,kriegnerMagneticAnisotropyAntiferromagnetic2017,szuszkiewiczSpinwaveMeasurementsHexagonal2006}. In this context, the two candidate magnetic point groups for in-plane moments are $m'm'm$ and $mmm$, corresponding to moments along the NNN and NN directions, respectively \cite{aoyamaPiezomag2024}.

In previous work on the orthorhombic antiferromagnetically ordered iron pnictide BaFe$_2$As$_2$ with twinned structural and magnetic domains below $T_{\mathrm{N}}$ \cite{doi:10.1126/science.1190482,dai_antiferromagnetic_2015}, a $\sim$10--20 MPa uniaxial compressive stress induces a small lattice distortion ($\Delta a/a\approx -1\times 10^{-3}$ near $T_{\mathrm N}\approx 140$ K, where $a$ is the orthorhombic lattice parameter) that detwins the structural and magnetic domains, but that is insufficient to change the direction of the ordered moment relative to the lattice \cite{PhysRevB.93.134519,liu_-plane_2020-1}.   
For $\alpha$-MnTe, despite a sizable spontaneous magneto-volume contraction below $T_{\mathrm{AM}}$ ($\left| \Delta V/V\right|>7\times 10^{-3}$) indicating a strong magnetoelastic coupling \cite{https://doi.org/10.1002/adfm.202305247}, the compressive strain applied in our neutron experiment is comparably small. From the load and cell geometry we estimate an upper bound $\Delta a/a \approx -1\times10^{-3}$ (i.e., $\epsilon \approx -0.1\,\%$), which is comparable to that in BaFe$_2$As$_2$ and lies below the resolution of conventional neutron diffraction measurements (Extended Data Fig.~\ref{figS:Braggstrain}).
 
In principle, a uniaxial compressive strain lowers the in-plane threefold rotational symmetry and can redistribute the in-plane magnetic domains observable in neutron diffraction pattern.
If this strain detwins the three 120$^\circ$ separated magnetic domains in $\alpha$-MnTe below $T_{\mathrm{AM}}$ -- thereby eliminating the domain averaging effect -- then neutron scattering experiments across $T_{\mathrm{AM}}$ should be able to conclusively identify the in-plane moment direction relative to the lattice (Figs. \ref{fig:fig1}e,g,i,k,m,o). Guided by this idea, we performed neutron diffraction measurements under two distinct strain geometries: compressive uniaxial strain applied along the NN Mn--Mn bond direction (Figs. \ref{fig:fig1}p,q,r) and, in a separate experiment, along the NNN Mn--Mn bond direction (Figs. \ref{fig:fig1}s, see Methods). We focus on the $(H,-0.5H) \times (0,K)$ plane with $L=1$ (r.l.u.), where six reciprocal-lattice positions with vanishing nuclear structure factor (marked in Fig.~\ref{fig:fig1}p) show magnetic intensity at $T=240$ K ($<T_{\mathrm{AM}}$) and no intensity at $T=320$ K ($>T_{\mathrm{AM}}$), confirming their magnetic origin. Under either strain direction, these six magnetic peaks separate into two symmetry-equivalent sets, marked by red and blue in Fig.~\ref{fig:fig1}t,u. Because peaks within each set transform identically, while the two sets respond differently as domains are redistributed as a function of increasing strain, we track the domain evolution using the ratio $I_{\mathrm{red}}/I_{\mathrm{blue}}$, where $I$ is the average integrated intensities of the red and blue set.

When compressive strain is applied along the NN Mn--Mn bond direction (Fig.~\ref{fig:fig1}r), the two red reflections, i.e., $(2,\bar{1},1)$ and $(\bar{2},1,1)$, are strongly enhanced while the other four blue reflections are suppressed, giving $I_{\mathrm{red}}/I_{\mathrm{blue}} = 3.190(8):1$. The observed ratio only matches the nearly single-domain $\mathrm{A}$ state within the $m^\prime m^\prime m$ magnetic point group (Fig.~\ref{fig:fig1}f,g). No domain configuration belonging to the alternative $mmm$ point group can reproduce this ratio, as demonstrated in Fig.~\ref{fig:fig1}t,u.
In contrast, applying strain along the NNN Mn--Mn bond direction produces the opposite response. As shown in Fig.~\ref{fig:fig1}s, the intensity pattern reverses, yielding $I_{\mathrm{red}}/I_{\mathrm{blue}} = 1:2.89(2)$. Although this ratio is compatible either with a 1:1 B/C domain pair within $m^\prime m^\prime m$ or a partially detwinned state within $mmm$ (Fig.~\ref{fig:fig1}t,u), our strain-dependence measurements show that it remains unchanged as the NNN strain is increased (Extended Data Fig.~\ref{figS:neutron_NNN}). The absence of strain evolution indicates that the domain population is already saturated as in Fig.~\ref{fig:fig1}h and thereby rules out the $mmm$ scenario in NNN strain geometry as in Fig.~\ref{fig:fig1}l,n. Therefore, while compressive strain applied along the NN direction gives a single-domain state, compressive strain applied along the NNN direction results in a two-domain state.

Taken together, the NN- and NNN-strain experiments independently converge to the same magnetic symmetry in $\alpha$-MnTe: the $m^\prime m^\prime m$ point group, with ordered moments lie along the NNN Mn--Mn bond direction regardless of the applied strain direction (Fig.~\ref{fig:fig1}f,h) \cite{aoyamaPiezomag2024}. Therefore, our results rule out strain directional dependent in-plane moment rotation in $\alpha$-MnTe, consistent with the notion that the applied strain is insufficient to modify the underlying easy axes dictated by SOC. We also demonstrate that, in free-standing sample, both magnetic point groups reproduce the magnetic intensity equally well due to domain averaging in line with previous neutron diffraction results (Figs. \ref{fig:fig1}d,e,j,k) \cite{kunitomiNeutronDiffractionStudy1964,efremdsaLowtemperatureNeutronDiffraction2005,szuszkiewiczNeutronScatteringStudy2005,kriegnerMagneticAnisotropyAntiferromagnetic2017,szuszkiewiczSpinwaveMeasurementsHexagonal2006}.

\section{Strain-tunable AHE and $T$-symmetry breaking}
Since $\alpha$-MnTe is an intrinsically hole-doped semiconductor, where holes arise from defects in the material's crystal structure \cite{IWANOWSKI201013},
we first systematically study the sample-dependent AHE in free-standing $\alpha$-MnTe single crystals. Figure \ref{fig:fig2}a shows the experimental setup where the externally applied magnetic field is along the $c$-axis, and the three 120$^\circ$ separated magnetic domains are shown in the inset of Fig.~\ref{fig:fig2}a and in Fig.~\ref{fig:fig2}b. Fourteen samples were measured, and the results are summarized in Figs. \ref{fig:fig2}d-f. Although all measured samples have a similar $T_{\mathrm{AM}}$ (Fig.~\ref{fig:fig2}d), the spontaneous AHE is only observed for samples with charge gaps falling between 15--18 meV (Figs. \ref{fig:fig2}e,f). The inset of Figure \ref{fig:fig2}c shows the representative field-dependence of the Hall resistivity/magnetization with linear background subtracted (see Methods for background subtraction details), which exhibits clear hysteresis loops. The main panel of Figure \ref{fig:fig2}c shows the temperature dependence of the anomalous Hall resistivity ($\rho_{xy}^{\mathrm A}$, defined as $\rho_{xy}$ at zero field extracted from positive-to-negative field sweeps), which was only observed between 200--300 K, consistent with the earlier work on free-standing bulk $\alpha$-MnTe \cite{kluczykCoexistenceAnomalousHall2024}.

To determine the strain-dependent AHE, we applied tunable uniaxial strain along both the NNN and NN Mn--Mn bond directions in $\alpha$-MnTe single crystals exhibiting the AHE.

The total uniaxial strain experienced by the sample, $\epsilon_{\mathrm{tot}}$, has two contributions: a nominal piezo-driven mechanical strain ($\epsilon_{\mathrm{nom}}$) and a thermal-mismatch strain ($\epsilon_{\mathrm{th}}$), arising from the different thermal expansion coefficients of MnTe and the titanium frame of the strain cell ($\epsilon_{\mathrm{th}}\approx0.2\,\%$ at 200~K; see Methods and Extended Data Fig.~\ref{figS:Silicon}). Throughout the manuscript, the displayed strain values correspond to the total effective sample strain,
$\epsilon_{\mathrm{tot}}=\mu\left(\epsilon_{\mathrm{th}}+\epsilon_{\mathrm{nom}}\right),$ where $\eta$ is the strain-transmission ratio (see Methods and Extended Data Fig.~\ref{figS:FEA}).

We first focus on the NNN configuration. Figure~\ref{fig:fig3}a shows 
the experimental setup. The dark and light gray lines in Fig.~\ref{fig:fig3}b show the hysteresis loop of the background subtracted Hall resistivity $\Delta\rho_{xy}$ in a strain-free 
$\alpha$-MnTe sample at 290 K, measured before mounting on the piezo-strain cell. We note that the hysteresis loop in the strain-free sample is always highly stretched, indicating a high level of disorder \cite{aminNanoscaleImagingControl2024}. However, when sample experiences a tensile uniaxial strain ($\epsilon_{\mathrm{tot}}\approx0.11\,\%$), the switching of anomalous Hall resistivity sharpens and the coercive fields ($H_c^{\pm}$) are significantly reduced, forming a square-like hysteresis loop (more details in Extended Data Fig.~\ref{figS:AHEsubtraction}). Such tensile strain along the NNN direction is symmetry-equivalent to a compressive uniaxial strain along the NN Mn--Mn bond direction, which stabilizes a single magnetic domain according to our neutron scattering analysis (Fig.~\ref{fig:fig1}f). This dramatic change suggests that the formation of a single in-plane magnetic domain and 
the elimination of domain boundaries 
is critical to the optimal control of the $\rightleftarrows$ and $\leftrightarrows$ altermagnetic domains related by time reversal (Figs. \ref{fig:fig1}d,e).  

Figures \ref{fig:fig3}c-e show the field dependent $\Delta\rho_{xy}$ under various compressive and tensile strains at $T=240, 233,$ and 230 K, respectively. The corresponding longitudinal resistivity $\Delta\rho_{xx}$ is shown in Figs. \ref{fig:fig3}f-h.  At 240 K,  the size of AHE reduces systematically  without significant change of $H_c(T)$ when the strain is tuned from compressive to tensile.
On decreasing temperatures to 233 K (Fig.~\ref{fig:fig3}d) and 230 K (Fig.~\ref{fig:fig3}e), we continue to see a strong tunability of AHE by strain without much change in 
$H_c(T)$.  Surprisingly, at $T=233$ K the sign of the AHE switches from negative to positive when the strain is tuned from compressive to tensile. At $T=230$ K, the sign of the AHE becomes positive and its size increases when the strain is tuned toward the tensile side. A nearly identical sign-switching AHE controlled by uniaxial strain along the NN bond direction is shown in Extended Data Fig.~\ref{figS:nnsignchange}.
Since the size of the coercive field remains about the same when the AHE vanishes, which can be seen from the longitudinal resistivity $\rho_{xx}$ (Fig.~\ref{fig:fig3}g and Extended Data Fig.~\ref{figS:nnsignchange}b), we conclude that the weak $c$-axis canted moment remains finite throughout the sign change of AHE. Therefore, this phenomenon cannot be explained by the piezomagnetic effect, as has been observed in Mn$_3$Sn where the net magnetization reverses while the anomalous Hall vector remains unchanged \cite{ikhlas2022piezomagnetic,aoyamaPiezomag2024,beyUnexpectedTuningAnomalous2024}. Instead, it must arise from strain-induced changes in the electronic $\mathbf{k}$-dependent Berry curvature \cite{Takahashi2025,gonzalezbetancourtSpontaneousAnomalousHall2023,kluczykCoexistenceAnomalousHall2024}. Finally, Figure \ref{fig:fig3}i summarizes the phase diagram of the strain-dependent AHE,  $\rho_{xy}^{\rm A}(\epsilon_{\mathrm{tot}})$.

\section{Temperature and strain-tunable AHE}
\begin{figure*}[ht!]
    \includegraphics[width = 180mm]{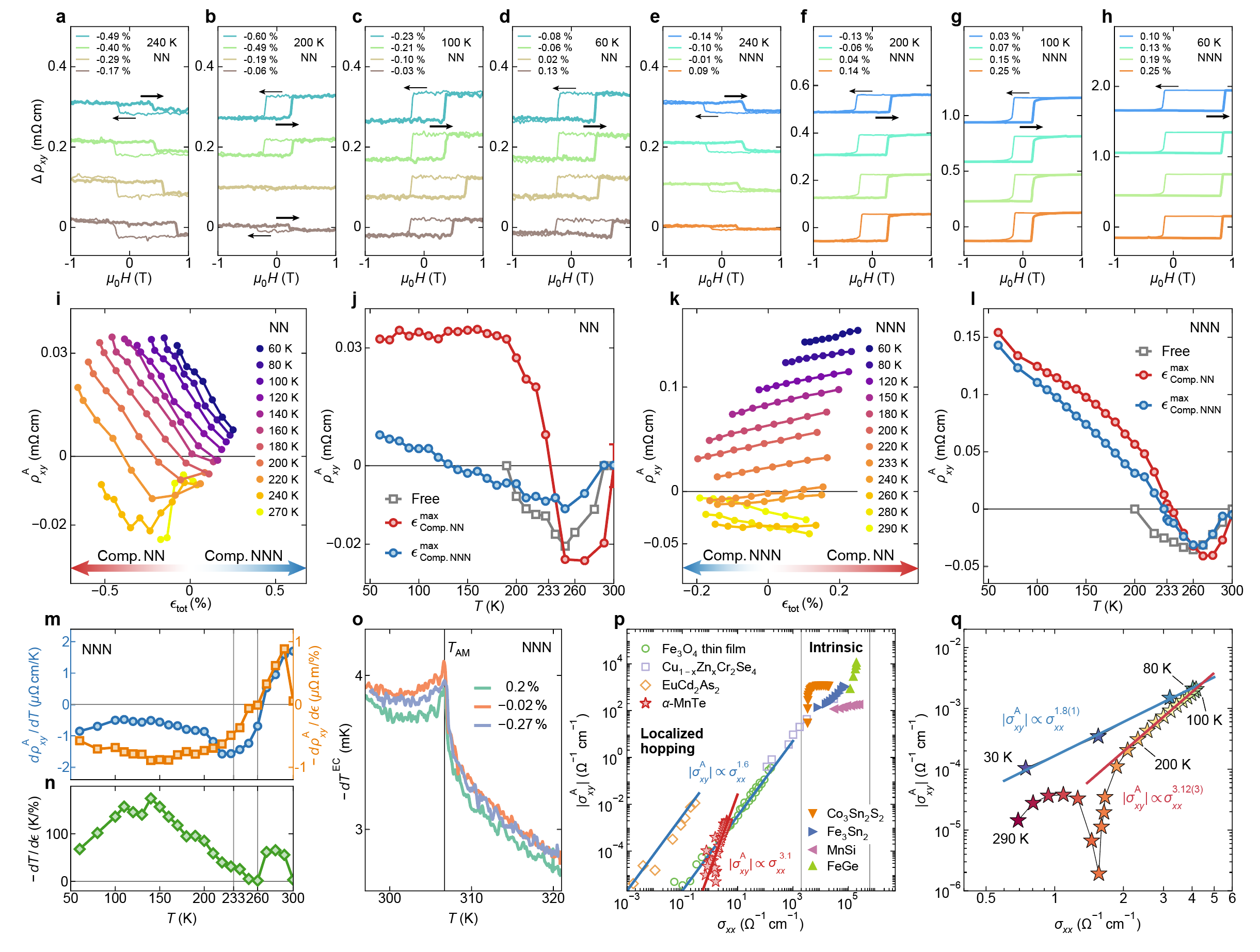}
    \caption{
    \textbf{Strain tunable AHE across all temperatures.}
    \textbf{a-d,} Hall resistivity $\Delta\rho_{xy}$ (linear background subtracted) measured at various temperatures under different strain levels with uniaxial stress $\sigma$ applied along the NN Mn--Mn bond direction. The magnetic-field sweep directions follow the definition in Fig.~\ref{fig:fig3}. 
    \textbf{e-h,} Corresponding Hall resistivity $\Delta\rho_{xy}$ with uniaxial stress $\sigma$ applied along the NNN Mn--Mn bond direction. The hysteresis loops exhibit a sign reversal below 233 K. A pronounced exchange-bias effect is evident in \textbf{h}. Much weaker effects also occur at higher temperatures and can be modified by changing the magnetic-field sweep range. To ensure consistent comparison across temperatures and strain levels, all sweeps were performed up to $\pm 9$\,T. The strain values in the legends of \textbf{a--h} denote the total strain experienced by the sample, $\epsilon_{\mathrm{tot}}$.
    \textbf{i,} $\rho_{xy}^{\mathrm A}$ as a function of total NN strain $\epsilon_{\mathrm{tot}}$ at different temperatures. Arrows indicate the mapping between strain sign and equivalent compressive loading: positive NN strain (tension along NN) is symmetry-equivalent to compression along the NNN direction (blue arrow), whereas negative NN strain corresponds to compression along NN (red arrow); see the corresponding neutron results in Fig.~\ref{fig:fig1}p,r,s.
    \textbf{j,} Temperature dependence of $\rho_{xy}^{\mathrm A}$ evaluated at the two strain extrema in \textbf{i}: most negative NN-strain (maximum NN-compressive strain $\epsilon_{\mathrm{Comp.\,NN}}^{\mathrm{max}}$), and most positive NN-strain (symmetry-equivalent to maximun NNN-compressive strain $\epsilon_{\mathrm{Comp.\,NNN}}^{\mathrm{max}}$ loading), compared with the free-standing data (grey).
    \textbf{k,l,} Analogous to \textbf{i,j}, but for the NNN-loading configuration.
    \textbf{m,} Temperature dependence of $d\rho_{xy}^{\mathrm A}/dT$ and $d\rho_{xy}^{\mathrm A}/d\epsilon$ with strain along the NNN direction. The vertical lines indicate 233 K and 260 K, corresponding to the temperatures where $\rho^{\mathrm A}_{xy}$ crosses zero and reaches its maximum value, respectively.
    \textbf{n}, $-dT/d\epsilon$ versus temperature, calculated as $-(d\rho_{xy}^{\mathrm A}/d\epsilon)/(d\rho_{xy}^{\mathrm A}/dT)$.
    \textbf{o,} Elastocaloric oscillation temperature $-dT^{\mathrm EC}$ at various strain levels near the $T_{\mathrm{AM}}$, showing no evident shift in $T_{\mathrm{AM}}$ under strain.
    \textbf{p,} Log-log plot of $|\sigma_{xy}^{\mathrm A}|$ versus $\sigma_{xx}$ in various magnetic materials \cite{liu_giant_2018,ye_massive_2018,lee_hidden_2007, tengDiscoveryChargeDensity2022a,klemm_vacancy-induced_2025,fernandez2008universal,PhysRevLett.99.086602,shiAbsenceWeylNodes2024a}. The blue lines represent the typical scaling behavior in semiconductors, with an exponent of 1.6 in the localized hopping regime. The AHC (anomalous Hall conductivity) in $\alpha$-MnTe is dramatically different from others.
    \textbf{q,} The zoom-in AHC in $\alpha$-MnTe. It exhibits two primary scaling behaviors. A scaling exponent of $\sim$ 1.8(1) appears below 80 K (blue line), where the resistivity diverges, consistent with localized hopping picture. In contrast, the scaling exponent is $\sim$ 3.12(3) between 100--200 K (red line). }
    \label{fig:fig4}
\end{figure*}

Figures \ref{fig:fig4}a-h show the AHE hysteresis loops measured at different temperatures under uniaxial strain applied along the NN and NNN Mn--Mn bond directions. In the single domain states - stabilized naturally by $\epsilon_{\mathrm{th}}$ in the NNN configuration or by applying sufficiently large compressive uniaxial stress in the NN configuration (Fig.~\ref{fig:fig1}f) - the AHE hysteresis loops remains sharp and squared-liked at all temperatures, in stark contrast to the behavior of free-standing samples~\cite{kluczykCoexistenceAnomalousHall2024}.
Figures \ref{fig:fig4}i and k show the anomalous Hall resistivity $\rho_{xy}^\mathrm{A}$ as a function of strain applied along the NN and NNN directions at various temperatures, respectively.

Figures~\ref{fig:fig4}j,l present $\rho_{xy}^{\mathrm A}(T)$ at the two strain extrema that are symmetry-equivalent to compressive NN ($\epsilon^{\max}_{\mathrm{Comp.\, NN}}$, red) and compressive NNN ($\epsilon^{\max}_{\mathrm{Comp.\, NNN}}$, blue), together with $\rho_{xy}^\mathrm{A}$ measured without applying strain (grey). The extrema are labeled by their equivalent compressive direction to facilitate comparison with the neutron scattering results.

Compared with the strain-free sample \cite{kluczykCoexistenceAnomalousHall2024}, the detwinned single domain state (Figs. \ref{fig:fig1}f,g) exhibits AHE over a much wider temperature range consistent with a reduced density of domain boundaries. Remarkably, the AHE changes sign around 230 K, an energy scale comparable to the size of the 15--18 meV charge gap identified in Fig.~\ref{fig:fig2}f. 

We also note that applying NNN-strain from compressive to tensile on top of the existing $\epsilon_{\mathrm{th}}$ effectively translates the $\rho_{xy}^\mathrm{A}$ versus $T$ curve along the temperature axis. To illustrate this point, we plot [$d\rho_{xy}^\mathrm{A}/ dT$] and [$d\rho_{xy}^\mathrm{A}/ d\epsilon$] in Fig.~\ref{fig:fig4}m, which exhibit strikingly similar temperature dependence. By comparing these two curves, we conclude that the effect of 1$\,\%$ uniaxial strain is equivalent to a shift of AHE by $\sim$ 100 K over a wide temperature range (Fig.~\ref{fig:fig4}n).
A natrual explaination for the strain-induced temperature shift of the AHE is that uniaxial strain changes the $T_{\mathrm{AM}}$. To test this, 
we carried out strain and temperature-dependent elastocaloric effect measurements \cite{ikedaACElastocaloricEffect2019,PhysRevX.14.031015}, which track the change of entropy induced by strain, and are especially sensitive to the heat capacity anomaly near a phase transition. 
The altermagnetic phase transition is 
clearly visible as a peak in the elastocaloric effect at $T_{\mathrm{AM}}\approx 307$ K (Fig.~\ref{fig:fig4}o) \cite{kunitomiNeutronDiffractionStudy1964,efremdsaLowtemperatureNeutronDiffraction2005,szuszkiewiczNeutronScatteringStudy2005,kriegnerMagneticAnisotropyAntiferromagnetic2017,szuszkiewiczSpinwaveMeasurementsHexagonal2006}.
However, varying the applied uniaxial strain from compressive to tensile results in no measurable change of $T_{\mathrm{AM}}$, as observed in both the elastocaloric effect (Fig.~\ref{fig:fig4}o) and the derivative of resistance (Extended Data Fig.~\ref{figS:dRdT}). This robustness contrasts with the clear increase in transition temperature observed under a large hydrostatic pressure ($>0.3$ GPa)~\cite{sugiura1979effect, carlisle2025tuning}.

To gain more insight into the microscopic origin of the AHE in $\alpha$-MnTe,
we plot the anomalous Hall conductivity (AHC) $|\sigma_{xy}^\mathrm{A}|$ as a function of longitudinal conductivity $\sigma_{xx}$ on a log-log scale with other magnetic materials (Fig.~\ref{fig:fig4}p) \cite{liu_giant_2018,ye_massive_2018,lee_hidden_2007,tengDiscoveryChargeDensity2022a,klemm_vacancy-induced_2025,fernandez2008universal,PhysRevLett.99.086602,shiAbsenceWeylNodes2024a}. The $\sigma_{xx}$ of $\alpha$-MnTe is much smaller than all the metallic or semi-metallic magnetic materials in the intrinsic regime \cite{liu_giant_2018,ye_massive_2018,lee_hidden_2007,tengDiscoveryChargeDensity2022a,klemm_vacancy-induced_2025}. The low conductivity and short mean free path ($< 0.5$ nm, Extended Data Fig.~\ref{figS:RH_n}) place the observed AHE in $\alpha$-MnTe deep inside the localized hopping regime \cite{fernandez2008universal,PhysRevLett.99.086602,shiAbsenceWeylNodes2024a}.Figure \ref{fig:fig4}q shows a zoom-in of the $|\sigma_{xy}^\mathrm{A}|$ versus $\sigma_{xx}$ of $\alpha$-MnTe on a log-log scale. In the localized hopping regime, it was empirically observed that $\sigma_{xy}^\mathrm{A}$ is related to $\sigma_{xx}$ via $\sigma_{xy}^\mathrm{A}\propto\sigma_{xx}^{1.4\sim1.8}$, which has also been confirmed by theoretical calculations \cite{burkovAnomalousHallEffect2003a,liuScalingAnomalousHall2011a}. Below 80 K, where resistivity diverges (Fig.~\ref{fig:fig2}d), we indeed observed a scaling behavior (blue solid line) with an exponent of 1.8, consistent with localized hopping behavior. However, the AHC in $\alpha$-MnTe above 100 K deviates from this simple scaling law. Above 220 K, the AHC shows nonlinear behavior without a clear scaling relation. Most interestingly, in between 100 and 210 K, the AHC exhibits a perfect scaling behavior with an exponent of $\sim$ 3.1 (red solid line). This scaling exponent, which is tunable by strain (see Extended Data Figs.~\ref{figS:AHCscaling}a and b for the NN and NNN-strained samples, respectively), is beyond our current understanding of AHC \cite{RevModPhys.82.1539,osin2025extrinsic}.

\section{Summary and discussion}
Figures 1--4 reveal several important findings in bulk single crystals of  $\alpha$-MnTe: (i) A uniaxial compressive strain along the NN and NNN Mn--Mn bond directions effectively stabilize one (Fig.~\ref{fig:fig1}f) and two (Fig.~\ref{fig:fig1}h) in-plane magnetic domains, respectively. For both strain directions, the in-plane moment direction is along the NNN Mn--Mn bond direction relative to the underlying lattice
(Fig.~\ref{fig:fig1}b,f,h); (ii) The spontaneous AHE only appears in samples with charge gaps around 15--18 meV (Fig.~\ref{fig:fig2}f); (iii) Detwinning the sample into a single in-plane domain state sharpens the AHE hysteresis loop; (iv) All samples exhibit a dominant linear dependence of the AHE on uniaxial strain; (v) The AHE reverses its sign both as a function of temperature and strain at $\sim$230 K or $\sim$20 meV (Fig.~\ref{fig:fig3}i). The AHE in the single domain state also persists down to a much lower temperature range compared to free standing sample (Figs. \ref{fig:fig4}j,l); 
(vi) There is no observable change in $T_{\mathrm{AM}}$ or a new magnetic transition induced under applied uniaxial NNN compressive strain up to $\epsilon=-0.27\,\%$ and tensile strain up to $\epsilon=0.2\,\%$ (Fig.~\ref{fig:fig4}o); (vii) The spontaneous AHC in $\alpha$-MnTe falls within the localized hopping regime at low temperatures, yet it exhibits an unusual scaling exponent in the intermediate temperature range (Figs. \ref{fig:fig4}p,q and Extended Data Fig.\ref{figS:AHCscaling}).

To understand the microscopic origin of these observations, we consider the following picture. A free-standing (strain-free) $\alpha$-MnTe has
three 120$^\circ$ separated in-plane magnetic domains (Fig.~\ref{fig:fig1}d) and two $\rightleftarrows$ and $\leftrightarrows$ magnetic stacked domains associated with each in-plane moment direction (Fig.~\ref{fig:fig1}a). While a $c$-axis aligned magnetic field in the range of a few Tesla in the AHE measurements (Fig.~\ref{fig:fig2}) cannot modify the 
population of the three in-plane domains, it switches the two $\rightleftarrows$ and $\leftrightarrows$ domains via the coupling to the weak $c$-axis FM moment \cite{aminNanoscaleImagingControl2024}, which gives rise to the observed AHE hysteresis  \cite{gonzalezbetancourtSpontaneousAnomalousHall2023,kluczykCoexistenceAnomalousHall2024}. However, the AHE in the strain-free $\alpha$-MnTe can only be observed in the temperature regime above $\sim$180 K and below $T_{\mathrm{AM}}$. The hysteresis loop is also highly stretched, indicative of high level of magnetic disorder. The magnetic disorder is likely originated from the in-plane magnetic domain boundaries, at which the in-plane moment direction does not allow the existence of canted moment and AHE. When applying tensile strain along the NNN direction or equivalently compressive strain along the NN direction, it induces a single in-plane magnetic domain and eliminates domain boundaries (Fig.~\ref{fig:fig1}f), thus sharpening the hysteresis loop and reducing the coercive field $H_c$ (Fig.~\ref{fig:fig3}b). The temperature range where the AHE can be observed is also significantly extended. 

It is important to contrast our results with previous work on Mn$_3$\textit{X} (\textit{X} = Sn, Ge, Pt), which reported a sign change of the AHE with in-plane uniaxial strain or hydrostatic pressure \cite{ikhlas2022piezomagnetic,Zuniga-Cespedes_2023,sukhanov2018gradual,reis2020pressure,singh2020pressure}. First, while these compounds display a non-collinear magnetic configuration that hosts a multipolar moment similar to $d$-wave altermagnets, they are not altermagnets \cite{WOS:001147635900001,Jungwirth2025symmetry}. In contrast, the collinear magnetic configuration of $\alpha$-MnTe corresponds to a $g$-wave altermagnet, whose symmetries are therefore distinct from both $d$-wave altermagnets and Mn$_3$\textit{X}. 
Second, Mn$_3$Sn has a giant piezomagnetic coefficient, which makes its net magnetization, which contributes to the AHE, very sensitive to strain. Specifically, at the critical strain value for which the sign-change of the AHE is observed in Mn$_3$Sn, a concomitant reversal of the net magnetization also takes place \cite{ikhlas2022piezomagnetic}, strongly supporting a scenario in which the strain-induced changes in the net magnetization are responsible for the strain-induced changes in the AHE. In contrast, using the reported piezomagnetic coefficient of  $\alpha$-MnTe  \cite{aoyamaPiezomag2024}, and estimating the stress applied in our sample to be 100 MPa, we obtain an induced FM moment of about $10^{-6} \mu_B$, much smaller than the measured FM moment of about $10^{-5} \mu_B$ of our unstrained sample [see inset of Fig.~\ref{fig:fig2}c]. This is consistent with our elastocaloric observation that the largest strains used in our experiment are insufficient to modify $T_{\mathrm{AM}}$. Third, in Mn$_3$Sn, the Hall vector associated with the observed AHE is in the same plane of the non-collinear magnetic moments, whereas in our case it is perpendicular to the collinear moments.

The considerations above imply that piezomagnetism is not the origin of the observed strain dependence of the AHE in  $\alpha$-MnTe, and that a different mechanism than the strain-induced changes in the AHE of Mn$_3$Sn is at play. Moreover, while a strain-induced change in the altermagnetic ground state was theoretically proposed \cite{belashchenkoGiantStrainInducedSpin2025}, the required strain values are much larger than the ones applied in our experiments.  We propose, instead, that it arises from strain-induced changes in the $z$-component of the electronic Berry curvature $\Omega_z(\textbf{k})$. Without SOC and in the absence of applied strain, $\Omega_z(\textbf{k})$ is a Berry curvature hexadecapole, consistent with the $g$-wave symmetry of the spin-splitting of the bands in $\alpha$-MnTe \cite{krempaskyAltermagneticLiftingKramers2024,aminNanoscaleImagingControl2024,PhysRevLett.133.156702,gonzalezbetancourtSpontaneousAnomalousHall2023,kluczykCoexistenceAnomalousHall2024,leeBrokenKramersDegeneracy2024,loveseyTemplatesMagneticSymmetry2023,mazinAltermagnetismMnTeOrigin2023,PhysRevLett.132.236701,PhysRevLett.133.086503,839n-rckn}. As such, it cannot give rise to a finite AHE $\sigma_{xy}$. Generally, however, SOC \cite{doi:10.1126/sciadv.aaz8809,kluczykCoexistenceAnomalousHall2024} and strain \cite{Takahashi2025} can distort the Berry curvature multipole of altermagnets, resulting in a net Berry curvature monopole and thus a finite $\sigma_{xy}$. We investigate both possibilities via a phenomenological model (see Methods). Without SOC, we find that in-plane strain alone cannot induce a non-zero $\sigma_{xy}$. In contrast, SOC alone, without strain, generates a non-zero AHE that depends on the two-component altermagnetic order parameter $\textbf{L}$ as $\sigma_{xy} = A L^3$ (see also \cite{Fernandes2024,Mcclarty2024}), which is consistent with our observations (here, $A$ is a temperature-dependent parameter). Interestingly, we find that the combination of SOC and uniaxial strain $\epsilon$ along the NNN Mn--Mn direction leads to a new contribution to the AHE,  $\sigma_{xy} = A L^3 + B\epsilon L$  ($B$ is another parameter), provided that the system is not in a three-domain state. Crucially, this contribution is not only linear in the strain, but also linear in the altermagnetic order parameter. As a result, close to the temperature where the zero-strain AHE switches sign, it is reasonable to expect that this additional contribution can become comparable to the zero-strain AHE, causing the observed sign change induced by strain. Indeed, we can fit the dependence of $\sigma_{xy}$ on $\epsilon$ in both cases of strain applied along the NNN and NN directions using this expression, as shown in Extended data Fig. \ref{figS:model_fit}. Performing this fitting over all temperatures enables us to extract the temperature-dependent combinations $C_0 \equiv A L^3$ and $C_1 \equiv BL$, plotted in Extended data Fig. \ref{figS:model_fit}. The non-monotonic behavior of both quantities indicates that the parameters $A$ and $B$ have their own temperature dependence, which ultimately reflects the microscopic mechanism by which an AHE appears in a self-doped semiconducting material. Nevertheless, consistent with our conclusions, we find that $A L^2 \ll B$ near the regimes in which we observe that even small strain values can change the sign of the AHE. Note that the magnitudes of the coefficients $C_0$ are rather different for the two samples in which NNN and NN strains were applied, reflecting the sensitivity of the AHE on sample properties observed in Fig. \ref{fig:fig2}f. Interestingly, this is not the case for the coefficients $C_1$, which show similar behaviors for the two samples -- the opposite signs just reflect the fact that tensile strain along the NNN direction is equivalent to compressive strain along the NN direction.

While a microscopic model to describe the Berry curvature of $\alpha$-MnTe is beyond the scope of this work, the strong dependence of the AHE on the charge gap, shown in Fig.~\ref{fig:fig2}f, suggests that the temperature dependence of the number of hole-like and electron-like carriers in the semiconductor plays a key role in enabling a non-zero AHE. An intrinsic origin of the AHE in $\alpha$-MnTe is also consistent with a recent theoretical analysis \cite{osin2025extrinsic}. In this regard, it is interesting to note that the unexpected temperature-induced sign change of the AHE in samples where the nominal strain is zero happens in the temperature range where $|\sigma_{xy}|$ does not show a clear scaling behavior with $\sigma_{xx}$, as shown in Fig.~\ref{fig:fig4}q. 
More broadly, our experiments demonstrate that detwinned material with a single-domain structure with moments along the NNN Mn--Mn bond direction in altermagnetic $\alpha$-MnTe  enables an abrupt switching of the AHE with reduced coercive field over a wide temperature range \cite{gonzalezbetancourtSpontaneousAnomalousHall2023,kluczykCoexistenceAnomalousHall2024}. Therefore, altermagnetic $\alpha$-MnTe can potentially be used as highly scalable, strain-tunable magnetic sensors and spintronic devices with vanishing fringing fields for near room temperature operation \cite{RevModPhys.76.323,RevModPhys.90.015005,PhysRevX.12.040501,WOS:000775805300001,daiResearchProgressFuture2025,fenderAltermagnetismChemicalPerspective2025,WOS:001147635900001,bangarInterplayAltermagneticOrder2025,PhysRevB.108.075425,Jungwirth_Fernandes2025,Jungwirth2025symmetry}.

\section*{Methods}\label{method}
\subsection{Sample preparation and characterizations}
Single crystals of $\alpha$-MnTe were synthesized using flux and chemical vapor transport (CVT) methods \cite{kluczykCoexistenceAnomalousHall2024}. The flux methods can produce centimeter-size single crystals (Extended Data Fig.~\ref{figS:characterization}a) while the CVT method typically produces crystals of size $\sim\!1\times1\times1$ mm$^3$. The phase purity of the crystals was confirmed by powder X-ray diffraction (Rigaku SmartLab II), followed by Rietveld refinement using the FullProf suite. Chemical composition was verified by energy-dispersive X-ray spectroscopy (EDS) in an FEI Nano 450 scanning electron microscope at the Shared Equipment Authority at Rice University (Extended Data Fig.~\ref{figS:characterization}). No significant differences were found across samples grown by different methods using these probes. The crystals used for the neutron and transport measurements in this study were obtained primarily from flux growth. 

The in-plane crystal axes were determined by X-ray Laue diffraction in a back-scattering geometry (Extended Data Fig.~\ref{figS:characterization}b), after which the samples were manually polished into rectangular plates with the long axis in the NNN Mn--Mn bond direction. The typical dimensions are about $1.5\times 1 \times 0.1$ mm$^{3}$. Electrical contacts in a standard five-terminal Hall bar geometry (Fig.~\ref{fig:fig2}a) were first fabricated by sputtering a $\sim$40 nm thick gold layer to reduce contact resistance. Then the contacts were connected to the sample holder using DuPont 4929N silver paste and gold wires (25 $\mu$m). Transport measurements were performed in a Quantum Design DynaCool system, with signals detected by SR860 lock-in amplifiers. High-field transport measurements were conducted in a resistive magnet ($\pm$35 T) at the National High Magnetic Field Laboratory in Tallahassee, Florida (Extended Data Fig.~\ref{figS:highfield}). 

Magnetic susceptibility and magnetization were measured in a Quantum Design MPMS system using SQUID-VSM mode (Extended Data Figs.~\ref{figS:characterization}d,e). The larger susceptibility for out-of-plane fields indicates that the $c$-axis is the easy axis. The peak feature in $d\chi/dT$ denotes the altermagnetic transition temperature $T_{\mathrm{AM}}$. These results are consistent with previous report \cite{kluczykCoexistenceAnomalousHall2024}. 

\subsection{Uniaxial strain transport experiments}
In-situ tunable uniaxial strain was applied to a thin-plate sample using a homemade three-piezostack uniaxial strain cell based on the technology in Refs. \cite{hicksPiezoelectricbasedApparatusStrain2014a,mutchEvidenceStraintunedTopological2019}. The polished sample was glued across the gap of the strain cell using Stycast 2850FT epoxy, with a typical gap size of $\sim$1 mm. In practice, a thin cigarette paper ($\sim$0.03 mm thick) was also inserted beneath the sample to prevent direct contact between the sample and the titanium plates. Uniaxial strain was applied to the sample by tuning the gap size, controlled by differential motion of the three piezostacks driven by an external DC voltage. In our experiments, we utilized a Keithley 2400 source meter to supply positive voltage on the inner piezostack and reverse voltage to the outer piezostacks to generate compressive strain, and the opposite voltage configuration was used to apply tensile strain.  The amplitude of the nominal strain $\epsilon_{\mathrm {nom}}$ can be extracted by measuring the strain on the piezostack $\epsilon_{\mathrm {piezo}}$ via a silicon strain gauge (SS-150–124–15P, Micron Instruments, Semi Valley, CA) glued to the side wall of one piezostack.  The strain on the piezostack is given by $\epsilon_{\mathrm {piezo}}=[(R-R_0)/R_0]/g$, where $R$ is the strain gauge resistance, $R_0$ is the resistance at zero strain, and $g$ is the gauge factor of the silicon strain gauge. The gauge factor is 80 at room temperature and 165 at 2 K, 
and the temperature dependence of $g$ was calibrated previously \cite{malinowskiSuppressionSuperconductivityAnisotropic2020}.
Consequently, the nominal strain can be obtained from: 
\begin{equation}
    \epsilon_{\mathrm {nom}} = 2\times\frac{\epsilon_{\mathrm {piezo}}\times L_{\mathrm {piezo}}/g}{L_{\mathrm {gap}}}
\end{equation}
where $L_{\mathrm {piezo}}$ and $L_{\mathrm {gap}}$ are the piezostack length (9 mm) and the gap size, respectively. In practice, however, the actual strain experienced by the sample often deviates from $\epsilon_{\mathrm {nom}}$ due to two major factors. The first is differential thermal expansion between the sample and the strain cell, which primarily shifts the effective zero-strain point. 
The thermal-expansion induced strain from the piezostacks is eliminated by the strain cell configuration \cite{hicksPiezoelectricbasedApparatusStrain2014a}. Thermal strain can nevertheless arise from differential contraction between the titanium frame and the sample, producing a temperature-dependent thermal strain $\epsilon_{\mathrm{th}}$ on the sample.
To quantify this effect, we estimate the temperature-dependent change of the frame gap using a silicon strain gauge. A silicon gauge was mounted on the strain cell, fixed at one end to measure the free-standing $R_0(T)$ and at both ends to measure $R(T)$ under thermal strain. As shown in Extended Data Fig.~\ref{figS:Silicon}a, $R(T)$ remains smaller than $R_0(T)$, indicating that the titanium gap decreases on cooling. Taking the free-standing measurement as the zero-strain baseline $R_0(T)$, we compute $\epsilon_{\mathrm{Si}}(T)$ (with gauge factor $g$ as calibrated previously) and convert it to the gap change via $\Delta \mathrm{Gap}(T)=\epsilon_{\mathrm{Si}}(T)\,L_{\mathrm{Si}}$.
Because the MnTe crystal also contracts on cooling, the net thermal strain experienced by a MnTe crystal glued across the gap is given by the mismatch between the frame-gap contraction and the MnTe in-plane lattice contraction \cite{baral2023giant}:
\begin{equation}
\epsilon_{\mathrm{th}}=\frac{\Delta \mathrm{Gap}}{\mathrm{Gap}}-\frac{\Delta a}{a}.
\end{equation}

Using the reported MnTe lattice parameter $a(T)$, we find that at 200~K the MnTe lattice contracts by $\sim$ 0.3$\,\%$ relative to room temperature, whereas the titanium gap contracts by only $\sim0.1\,\%$. The resulting mismatch corresponds to a tensile thermal strain of $\epsilon_{\mathrm{th}}\approx +0.2\,\%$ at 200~K.

The second factor is incomplete strain transfer through the epoxy layers, resulting in a transmission ratio $\mu<1$, which we estimate by finite element analysis (FEA). We performed FEA using the ANSYS Academic Research Mechanical software. The elastic properties of $\alpha$-MnTe were taken from the Materials Project database \cite{jain2013materials}, while those for the epoxy were provided by the manufacturer.
The FEA results are shown in Extended Data Fig.~\ref{figS:FEA}. In our simulation, we assumed epoxy layer with a total thickness of 50 $\mu$m, corresponding to the cigarette paper thickness (30 $\mu$m) plus the two thin epoxy layers (10 $\mu$m). The sample is enclosed by two epoxy layers of $\sim$ 50~$\mu$m each at the top and bottom ends, within a total gap size of 1 mm. By applying 1 $\mu$m displacement on the two ends, the nominal strain is $-0.2\,\%$. The FEA results in Extended Data Fig.~\ref{figS:FEA} exhibits a uniform strain distribution in the sample within the gap region, yielding a maximum strain of $-0.14\,\%$ at the center position. Consequently, by comparing to the nominal strain, it gives a strain transmission ratio of $\mu =$ 0.7. Accordingly, the total strain values displayed in the manuscript are given by $\epsilon_{\mathrm{tot}} = \mu(\epsilon_{\mathrm{nom}}+\epsilon_{\mathrm{th}})$, consistent with Ref. \cite{smolenski2025strain}.

The elastocaloric effect  measurements were performed using the same strain cell described above, but operated under both dynamic and static strains on different piezostacks \cite{ikedaACElastocaloricEffect2019,PhysRevX.14.031015}. The dynamic voltage supply was driven by a Stanford Research SR860 Lock-in amplifier following a TEGAM 2350 high-voltage amplifier. The oscillation temperature was monitored by a type-E thermocouple attached on the center of the sample using silver paste. The thermocouple signal was detected by the SR860, synchronized to the AC voltage source for piezostack. Considering the low tensile breaking point of hexagonal MnTe crystal at high temperatures, elastocaloric effect  measurements were only carried out in compressive strain regime by applying a positive DC voltage to the inner piezostack and oscillating AC voltage to the outer piezostacks. 

\subsection{Background subtraction of AHE}
\label{sec_AHEbkg_sub}
Hall resistivity $\rho_{xy}$ is often contaminated by the longitudinal $\rho_{xx}$ due to imperfect contacts alignment. However, as shown in Fig.~\ref{fig:fig3}b, the $\rho_{xy}$ of the free-standing sample differs from that of the strained sample. Accordingly, we employed two methods to subtract the linear background and extract the pure anomalous contribution to $\rho_{xy}$. For the free-standing data, the raw $\rho_{xy}$ data were first antisymmetrized as follows:
\begin{equation}
\begin{aligned}
    \rho_{xy, \mathrm{asym.}}^{B\downarrow} = (\rho_{xy}^{B\downarrow} - \rho_{xy}^{-B\uparrow})/2\\
    \rho_{xy, \mathrm{asym.}}^{B\uparrow} = (\rho_{xy}^{B\uparrow} - \rho_{xy}^{-B\downarrow})/2
\end{aligned}
\label{eq2}
\end{equation}
A linear background was then subtracted from both antisymmetrized datasets to obtain $\Delta\rho_{xy}^{B\downarrow}$ and $\Delta\rho_{xy}^{B\uparrow}$, respectively. The anomalous Hall resistivity shown in Fig.~\ref{fig:fig2}c and Fig.~\ref{fig:fig3}b is defined as: 
\begin{equation}
    \rho_{xy}^{\mathrm A} = \Delta\rho_{xy}^{B\downarrow=0}
    \label{eq3}
\end{equation}
A typical subtraction data is presented in Extended Data Fig.~\ref{figS:AHEsubtraction}a. 

In contrast to the spindle-shaped hysteresis loop of the free-standing sample, the AHE loop in the $\alpha$-MnTe sample under uniaxial strain exhibits a sharp transition at the coercive fields, as shown in Fig.~\ref{fig:fig3}b and Extended Data Fig.~\ref{figS:AHEsubtraction}. Furthermore, asymmetric coercive fields are observed, namely, exchange bias effect (Extended Data Fig.~\ref{figS:exchangebias}). Although the exchange bias $H_{eb}$ is weak at high temperatures (Extended Data Fig.~\ref{figS:exchangebias}d, inset), applying the simple antisymmetrization procedure of Eq.~\ref{eq2} introduces an artificial step at the coercive fields. To avoid this artifact, we employed an alternative method described below.

For example, the $\rho_{xy}^{B\downarrow}$ data (hereafter denoted simply as $\rho_{xy}^{\downarrow}$) can be divided into three magnetic-field windows: the left side of the step ($\rho_{xy}^{\downarrow \mathrm L}$), the step region ($\rho_{xy}^{\downarrow \mathrm M}$), and the right side of the step ($\rho_{xy}^{\downarrow \mathrm R}$). We first shift $\rho_{xy}^{\downarrow \mathrm L}$ vertically by removing a step size of $\delta\rho_{xy}$ to compensate $\rho_{xy}^{\downarrow \mathrm M}$ and obtain $\rho_{xy}^{\downarrow {\mathrm L}^\prime}$. Then we merge the $\rho_{xy}^{\downarrow {\mathrm L}^\prime}$ and $\rho_{xy\downarrow}^{\mathrm R}$ to form a new smooth dataset $\rho_{xy}^{\downarrow\mathrm{merg.}}$. This dataset is antisymmetrized according to 
\begin{equation}
    \rho_{xy}^{\downarrow\mathrm{as}} = [\rho_{xy}^{\downarrow\mathrm{merg.}}(+B) - \rho_{xy}^{\downarrow\mathrm{merg.}}(-B)]/2
\end{equation}
After antisymmetrization, $\rho_{xy}^{\downarrow\mathrm{as}}$ crosses zero field point. The final corrected dataset $\rho_{xy}^{\downarrow {\mathrm{C}}}$ is then reconstructed across the three magnetic-field windows as: 
\begin{equation}
\rho_{xy}^{\downarrow \mathrm{C}} =
\left\{
\begin{array}{ll}
\rho_{xy}^{\downarrow \mathrm{as,L}} + \delta \rho_{xy}/{2} & (\text{left region})\\[4pt]
\rho_{xy}^{\downarrow \mathrm{as,R}} - \delta \rho_{xy}/{2} & (\text{right region})
\end{array}
\right.
\end{equation}
The step region data in $\rho_{xy}^{\downarrow {\mathrm {C}}}$ was obtained by vertically shifting $\rho_{xy}^{\downarrow {\mathrm M}}$ with a proper constant.
The pure anomalous component $\Delta\rho_{xy}^{\downarrow}$ is then determined by subtracting a linear background, extracted by fitting $\rho_{xy}^{\downarrow \mathrm{as}}$ with a straight line. 
The same procedure is applied to the $\rho_{xy}^{B\uparrow}$ data to yield the anomalous component $\Delta\rho_{xy}^{\uparrow}$ (Fig.~\ref{fig:fig3}b–e, Extended Data Fig.~\ref{figS:AHEsubtraction}). Finally, the anomalous resistivity $\rho_{xy}^{\mathrm A}$ is obtained following the definition given in Eq.~\ref{eq3}.

\subsection{Neutron scattering experiments}
Neutron scattering under compressive uniaxial stress was performed at the Spallation Neutron Source, Oak Ridge National Laboratory (ORNL), USA. We performed initial experiments at SEQUOIA (BL-17), and follow-up measurements at CORELLI (BL-9) \cite{ye_implementation_2018-1} and HB-1A. The momentum transfer $\mathbf{Q}$ in three-dimensional reciprocal space in \(\mathrm{\AA}^{-1}\) was defined as $\mathbf{Q} = H\mathbf{a}^*+K\mathbf{b}^*+L\mathbf{c}^*$ where $H$, $K$, and $L$ are Miller indices and $\mathbf{a}^*=2\pi(\mathbf{b} \times \mathbf{c})/[\mathbf{a} \cdot (\mathbf{b} \times \mathbf{c})]$, $\mathbf{b}^*=2\pi(\mathbf{c} \times \mathbf{a})/[\mathbf{b} \cdot (\mathbf{c} \times \mathbf{a})]$, $\mathbf{c}^*=2\pi(\mathbf{a} \times \mathbf{b})/[\mathbf{c} \cdot (\mathbf{a} \times \mathbf{b})]$. For the hexagonal setting, we take \(\mathbf{a}=a\,\hat{\mathbf{x}}\), \(\mathbf{b}=a\left(\cos 120^{\circ}\,\hat{\mathbf{x}}+\sin 120^{\circ}\,\hat{\mathbf{y}}\right)\), and \(\mathbf{c}=c\,\hat{\mathbf{z}}\) with lattice parameters \(a = b\simeq 4.15~\mathrm{\AA}\) and \(c\simeq 6.72~\mathrm{\AA}\) at room temperature.
A square plate-shaped sample with mass of $\sim$ 90 mg ($\sim 4\times 4\times 0.5$ mm$^3$) was mounted inside a uniaxial strain device machined from 6061 aluminum alloy , as shown in Extended Data Fig.~\ref{figS:neutron_NN}a. The scattering plane in this configuration is ($H$, $K$, 0). In-plane uniaxial stress was applied by tightening the side M3$\times$0.5 screw. After the screw was in firm contact, an additional $1/4$-turn was applied to impose a uniaxial stress. The device was subsequently loaded into a cryofurnace capable of warming up to 400 K. Measurements in the strained configuration were performed at 320 K and then 240 K. The device was subsequently unloaded and the screw was loosened to remove stress. The sample was then reinstalled, and the same measurements were repeated at 320 K and 240 K. The neutron data were reduced and integrated by Mantid.

Since the current design lacks a loading spring, we estimate only an upper bound for the compressive strain applied to the sample. A $1/4$-turn of the M3$\times$0.5 screw would ideally produce a displacement of $\delta = 0.125$ mm. However, microscopic inspection indicates that the majority of the displacement was attributable to contact indentation at the aluminum plate-screw interface ($\sim$ 0.12 mm), leaving an effective displacement of $\sim$ 5 $\mu$m. Treating the aluminum frame and slider as rigid and allowing deformation only in the sample and the contacting aluminum plate, we model them as two springs connected in series with spring constants $k = E A/L$, where $E$ is the Young's modulus, $A$ is the cross-sectional area, and $L$ is the length along the stress direction. The spring constants of the sample and the aluminum plate are estimated to be 23.5 and 92 GPa$\cdot$mm, respectively, so the sample's share of the effective distortion is $\sim$ 4 $\mu$m, corresponding to a compressive strain of about $-0.1\,\%$. This strain value is comparable to our strain transport measurements. Nevertheless, this strain value is a rough estimate and should be regarded as an upper bound for the strain applied in our neutron experiments.

The applied uniaxial strain can have two major effects. As reported in previous strain studies in iron pnictides \cite{malinowskiSuppressionSuperconductivityAnisotropic2020}, a small uniaxial strain first reorients twin (nematic) domains and, once detwinning is achieved, can further tune the lattice constants and the associated order parameter. In $\alpha$-MnTe, a similar principle applies to magnetic domains: applying a uniaxial strain first detwins the magnetic domains, as evidenced by the changing of the hysteresis loop from a broad feature to sharp magnetic transitions at the coercive fields (Fig.~\ref{fig:fig3}b). Once a single-domain state is established, the uniaxial strain can further interact with altermagnetic order, as discussed later in the phenomenological model.

From a symmetry point of view, ideally, a compressive strain applied along either the NN or NNN Mn--Mn bond reduces the threefold rotational symmetry and imposes constraints on the magnetic domain configurations. For the magnetic structure in Fig.~\ref{fig:fig1}b, Domain $\mathrm{B}$ and $\mathrm{C}$ are symmetric about the strain axis and therefore their populations must remain equal ($r_\mathrm{B}=r_\mathrm{C}$) for any finite strain. The system can thus access a spectrum of states ranging from a single-domain state $\mathrm{A}$ (schematized in Fig.~\ref{fig:fig1}f) to a two-domain mixture $\mathrm{B}/\mathrm{C}$ in a $1{:}1$ ratio (Fig.~\ref{fig:fig1}h). For the alternative structure in Fig.~\ref{fig:fig1}c, the same symmetry consideration applies: Domain $\mathrm{B}^\prime$ and $\mathrm{C}^\prime$ remain symmetric about the strain axis, enforcing $r_{\mathrm{B}^\prime}=r_{\mathrm{C}^\prime}$ and allowing a corresponding continuum between the $\mathrm{A}'$ (Fig.~\ref{fig:fig1}l) and the $1{:}1$ mixture $\mathrm{B}'/\mathrm{C}'$ (Fig.~\ref{fig:fig1}n). These domain scenarios lead to different magnetic Bragg-peak patterns and can therefore be tested by neutron diffraction experiments.

In neutron scattering experiments, magnetic intensity at each Bragg peak is calculated as
\begin{equation}
I(\mathbf{Q})\propto\left|\hat{\mathbf{Q}}\times \mathbf{F}_\mathrm{M}(\mathbf{Q}) \times \hat{\mathbf{Q}}\right|^{2}
\label{eq6}
\end{equation}
where 
\begin{equation}
\mathbf{F}_\mathrm{M}(\mathbf{Q})=\sum_{j} f_j(Q)\,e^{-W_j}\,e^{i\mathbf{Q}\cdot\mathbf{r}_j}\,\bm{\mu}_j
\label{eq7}
\end{equation}
Here, $\mathbf{Q}$ is the scattering vector $(\hat{\mathbf{Q}}=\mathbf{Q}/|\mathbf{Q}|)$; $j$ labels the Mn ions in the unit cell; $f_j(Q)$ is the magnetic form factor; $\mathrm{e}^{-W_j}$ is the Debye--Waller factor; and $\mathbf{r}_j$ and $\boldsymbol{\mu}_j$ denotes the positions and moments of Mn, respectively. 
Equation~\ref{eq6} implies that only the component of the moment perpendicular to $\mathbf{Q}$ contributes to the magnetic Bragg intensity. Consequently, domains with larger $\boldsymbol{\mu}^{\perp \mathbf{Q}}$ components yield stronger intensity at that $\mathbf{Q}$. For example, in the free-standing sample (Fig.~\ref{fig:fig1}d and j), Domain $\mathrm{A}$ contributes unevenly to the six magnetic Bragg peaks (labeled in Fig.~\ref{fig:fig1}p) depending on the moment-$\mathbf{Q}$ angle, but the total intensity contributed from different domains at each peak remains the same due to the averaged domain populations. Once detwinned, in the single-domain $\mathrm{A}$ state (as shown in Fig.~\ref{fig:fig1}f,g), the $(2,\bar{1},1)$ and $(\bar{2},1,1)$ reflections (marked in red) are enhanced and the other four symmetry-related peaks (marked in blue) are suppressed. Nevertheless, the two-domain $\mathrm{B}/\mathrm{C}$ state leads to an opposite pattern: suppression of the  peaks and enhancement of the blue ones. In stark contrast, a single-domain $\mathrm{A}'$ state suppresses the  peaks, while the two-domain $\mathrm{B}'/\mathrm{C}'$ state enhances them.

Experimentally, as demonstrated in Fig.~\ref{fig:fig1}t,u, we extract the integrated intensities of two symmetry-related groups of magnetic Bragg reflections, indicated by the red and blue symbols. These two sets of peaks respond differently to the moment orientation for the domain configurations allowed under strain, and the resulting ratio $I_{\mathrm{red}}/I_{\mathrm{blue}}$ therefore serves as a observable to distinguish among possible domain scenarios for experimental data under NN and NNN strain.

For NN strain, as shown in Fig.~\ref{fig:fig1}r, the two reflections parallel to the strain axis, i.e.,$(2,\bar{1},1)$ and $(\bar{2},1,1)$, are enhanced while the other four are suppressed. Such an enhancement of the red peaks is, in principle, compatible with two distinct domain configurations: a single-domain $\mathrm{A}$ state or, alternatively, a two-domain $\mathrm{B}'/\mathrm{C}'$ state. However, the observed intensity ratio, $I_{\mathrm{red}}/I_{\mathrm{blue}} \sim 3.190(8):1$, can be reproduced only within the ABC domain manifold in a single domain $\mathrm{A}$ state; no $\mathrm{A}'\mathrm{B}'\mathrm{C}'$ combination yields a comparable ratio (see Fig.~\ref{fig:fig1}t,u).

For NNN strain, as demonstrated in Fig.~\ref{fig:fig1}s, the intensity pattern is reversed, giving $I_{\mathrm{red}}/I_{\mathrm{blue}} \sim 1:2.89(2)$. This suppression of the red peaks is consistent with two possible domain scenarios: a two-domain $\mathrm{B}/\mathrm{C}$ state or an intermadiate-detwinned state within the $\mathrm{A}'\mathrm{B}'\mathrm{C}'$ model, as indicated by the lower-right intersection in Fig.~\ref{fig:fig1}u. Based on the ratio alone, these two possibilities cannot be distinguished. The decisive evidence comes from our strain-dependence measurements: as the applied NNN strain is increased, the ratio $I_{\mathrm{red}}/I_{\mathrm{blue}}$ remains essentialy unchanged (Extended Data Fig.\ref{figS:neutron_NNN}). In the $\mathrm{A}'\mathrm{B}'\mathrm{C}'$ scenario, additional strain should continuously drive the system toward a more strongly detwinned $\mathrm{A}'$ state, producing a monotonic evolution of this ratio. The absence of such behavior therefore rules out the $\mathrm{A}'\mathrm{B}'\mathrm{C}'$ configuration.

Therefore, the combined NN- and NNN-strain measurements conclusively identify the $\mathrm{A}\mathrm{B}\mathrm{C}$ domain model as the correct description in both strain geometries, thus  
revealing that the in-plane moment direction
relative to the underlying lattice 
does not change with changing strain directions (Fig.~\ref{fig:fig1}b). 
Because uniaxial strain has a negligible effect on the ordered moment (see main text), we further infer that this spin configuration represents the intrinsic ground state of $\alpha$-MnTe.

We next perform a quantitative refinement of the magnetic Bragg intensities for the NN-strain configuration to demonstrate how the $\mathrm{A}\mathrm{B}\mathrm{C}$ model reproduces the measured pattern, as shown in Extended Data Fig.~\ref{figS:simulation}. The refinement details are discussed in the following paragraph.

In $\alpha$-MnTe, antiparallel stacking within the unit cell cancels magnetic scattering at even $L$, so magnetic intensity appears only at odd $L$. Accordingly, we restricted our refinements to magnetic Bragg peaks in the $L=\pm1$ planes (Extended Data Fig.~\ref{figS:simulation}) and, crucially, to nuclear-extinct positions (vanishing nuclear structure factor) so that the measured intensity is purely magnetic. We selected 12 reflections with identical $\lvert\mathbf{Q}\rvert$, thereby eliminating complications from the magnetic form factor and Debye--Waller factor. To refine for the domain population, the computed single-domain intensities $I_{\mathrm{A}}(H,K,L)$, $I_{\mathrm{B}}(H,K,L)$, and $I_{\mathrm{C}}(H,K,L)$ were combined as

\begin{equation}
\label{eq:Magrefine}
\begin{aligned}
I_{\mathrm{calc}}(H,K,L)
&= S
\big[
r_{\mathrm{A}} I_{\mathrm{A}}(H,K,L)\\
&+ r_{\mathrm{B}} I_{\mathrm{B}}(H,K,L)
+ r_{\mathrm{C}} I_{\mathrm{C}}(H,K,L)
\big].
\end{aligned}
\end{equation}

with a global scaling factor $S$ and domain volume fractions $r_{\mathrm{A}}, r_{\mathrm{B}}, r_{\mathrm{C}}\ge 0$ constrained by $r_{\mathrm{A}}+r_{\mathrm{B}}+r_{\mathrm{C}}=1$. The parameters $(S, r_{\mathrm{A}}, r_{\mathrm{B}}, r_{\mathrm{C}})$ were obtained by least-squares fits to the measured intensities. An analogous refinement was carried out for the alternative domain model $(\mathrm{A}',\mathrm{B}',\mathrm{C}')$ using the same procedure.

\subsection{Phenomenological model}
The paramagnetic phase of $\alpha$-MnTe is described by space group $P6_{3}/mmc$ (\#194). Without SOC, the altermagnetic order parameter can be expressed as the product of a $g$-wave form factor and a vector $\hat{\mathbf{n}}$ in spin space:
\begin{equation}
\mathbf{L}=g(\mathbf{k})\,\hat{\mathbf{n}}
\end{equation}
where $g(\mathbf{k})\propto k_{y}k_{z}\left(3k_{x}^{2}-k_{y}^{2}\right)$ transforms as the single-dimensional $\Gamma_{3}^{+}$ irreducible representation (irrep). The same symmetries imply that the Berry curvature (BC) $\boldsymbol{\Omega}\left(\mathbf{k}\right)$ transform as a hexadecapole:
\begin{equation}
\boldsymbol{\Omega}\left(\mathbf{k}\right)=g(\mathbf{k})\,\hat{\mathbf{n}}
\end{equation}
which, as a result, gives no AHE, since the
latter requires a net BC monopole. In principle, both SOC and strain
can distort a BC multipole into acquiring a net monopole moment \cite{Takahashi2025}. We first study these two effects separately and then combined.

We consider strain first, $\epsilon_{ij}=(\partial_{i}u_{j}+\partial_{j}u_{i})/2$,
where $\mathbf{u}$ is the lattice displacement vector. Using group
theory, the in-plane strain combination $\left(\epsilon_{x^{2}-y^{2}},-2\epsilon_{xy}\right)$ transforms as the two-dimensional $\Gamma_{5}^{+}$ irrep, whereas the out-of-plane shear strain combination $\left(\epsilon_{xz},\epsilon_{yz}\right)$ transforms as the two-dimensional $\Gamma_{6}^{+}$ irrep. As a result, there is no term that is linear in strain that can give a nonzero AHE. The leading order contribution is quadratic in strain and must involve out-of-plane shear strain
\begin{equation}
\sigma_{ab}\sim\left(\epsilon_{x^{2}-y^{2}}\epsilon_{yz}+2\epsilon_{xy}\epsilon_{xz}\right)n_{c}
\end{equation}
where $\sigma_{ab}$ is the antisymmetric conductivity tensor, $\left(abc\right)$
is a permutation of $(xyz)$, and $n_{c}$ is the direction of the magnetic moment (which, without SOC, is an arbitrary direction in spin space). We also find a cubic in strain contribution:
\begin{equation}
\sigma_{ab} \sim\epsilon_{yz}\left(3\epsilon_{xz}^{2}-\epsilon_{yz}^{2}\right)n_{c}
\end{equation}

These results show that pure in-plane strain alone cannot induce an AHE in $\alpha$-MnTe. We now proceed to consider the role of SOC, which makes $\hat{\mathbf{n}}$ must transform as space-group irreps. Specifically, $n_{z}$ transforms as $m\Gamma_{2}^{+}$ while the combination $\left(n_{y},\,-n_{x}\right)$ transforms as the two-dimensional $m\Gamma_{6}^{+}$ irrep. As a result, in the case of out-of-plane moments, the altermagnetic (AM) order parameter $L_{z}\propto g\left(\mathbf{k}\right)n_{z}$ transforms as $m\Gamma_{4}^{+}$. For in-plane moments, as relevant for $\alpha$-MnTe, the order parameter $\mathbf{L}=\left(L_{x},L_{y}\right)$ transforms as the $m\Gamma_{5}^{+}$ irrep, with $L_{x}\propto g\left(\mathbf{k}\right)n_{x}$ and $L_{y}\propto g\left(\mathbf{k}\right)n_{y}$. The Landau theory of this altermagnetic order parameter, $\mathbf{L}=L\left(\cos\theta,\,\sin\theta\right)$, corresponds to a six-state clock model \cite{Fernandes2024,Mcclarty2024,Chakraborty2025}:
\begin{equation}
F_{\mathrm{AM}}=\frac{a}{2}\,L^{2}+\frac{u}{4}\,L^{4}+\frac{\gamma}{6}\,L^{6}\cos6\theta
\end{equation}
Here, $a=(T-T_{AM})/T_{AM}$, $\gamma>0$ selects the six degenerate states $\theta=\frac{\left(2n+1\right)\pi}{6}$, with $n=0,\cdots,5$, whereas $\gamma<0$ selects the six degenerate states $\theta=\frac{n\pi}{3}$, with $n=0,\cdots,5$. Note that the configurations that are symmetry-related to $\mathbf{L}\propto(1,0)$, i.e. moments along the $x$-axis, correspond to $\theta=\frac{n\pi}{3}$ whereas the configurations that are symmetry-related to $\mathbf{L}\propto(0,1)$, i.e. moments along the $y$-axis, correspond to $\theta=\frac{\left(2n+1\right)\pi}{6}$. The latter is the one realized in $\alpha$-MnTe, as the moments point along the $[1,\bar{1},0]$ direction, which corresponds to the $y$-axis. The in-plane altermagnetic order parameter can generate both an AHE (irrep $m\Gamma_{2}^{+}$) and an orthorhombic lattice distortion (irrep $\Gamma_{5}^{+}$) according to \cite{Chakraborty2025}:
\begin{align}
\sigma_{xy} & \sim3L_{x}^{2}L_{y}-L_{y}^{3}=L^{3}\sin3\theta\label{eq:Mz}\\
\left(\begin{array}{c}
\epsilon_{x^{2}-y^{2}}\\
-2\epsilon_{xy}
\end{array}\right) & \sim\left(\begin{array}{c}
L_{x}^{2}-L_{y}^{2}\\
-2L_{x}L_{y}
\end{array}\right)=L^{2}\left(\begin{array}{c}
\cos2\theta\\
-\sin2\theta
\end{array}\right)\label{eq:epsilon}
\end{align}
Note that $\sigma_{xy}$ transforms as the same irrep as the out-of-plane magnetization $M_{z}$, so the same relationship holds for $M_{z}$. Importantly, while both sets of values of $\theta$ give a non-zero orthorhombic distortion, only the $\theta=\frac{\left(2n+1\right)\pi}{6}$ values give a non-zero $\sigma_{xy}$. Thus, an AHE only appears when the moments point along the $y$-axis (and symmetry-related directions), explaining why $\alpha$-MnTe displays an AHE even without applied strain \cite{Mcclarty2024}. 

We now proceed to investigate what happens when both SOC and strain are present. We consider application of in-plane uniaxial strain $\epsilon$ along the $\mathbf{a}$-axis direction that makes an angle $\alpha$ with respect to the $x$-axis, which corresponds to the $[1,1,0]$ direction. Using group theory, we find that the combination:
\begin{equation}
\boldsymbol{\epsilon}=\left(\begin{array}{c}
\epsilon_{x^{2}-y^{2}}\\
-2\epsilon_{xy}
\end{array}\right)=\epsilon\left(\begin{array}{c}
\cos2\alpha\\
-\sin2\alpha
\end{array}\right)
\end{equation}
transforms as the irrep $\Gamma_{5}^{+}$. We can thus obtain the Landau free-energy for the altermagnetic order parameter in the presence of strain (to linear order in strain):
\begin{align}
F'_{\mathrm{AM}} & =\left(\frac{a}{2}\,L^{2}+\frac{u}{4}\,L^{4}+\frac{\gamma}{6}\,L^{6}\cos6\theta\right)\label{eq:F_tot}\\
 & +\lambda_{1}L^{2}\epsilon\cos\left(2\theta-2\alpha\right)+\lambda_{2}L^{4}\epsilon\sin3\theta\sin\left(\theta+2\alpha\right)\nonumber 
\end{align}
Minimization of the free energy for a fixed $\epsilon$ and $\alpha$ gives the $\theta$ and $L$ values. While a full solution of this free-energy is beyond the scope of this work, we perform a qualitative analysis for the case $\alpha=\pi/2$, corresponding to the experimental situation of uniaxial strain along the $y$-axis (i.e., parallel to the moments direction or along the NNN Mn-Mn bond direction). For $\alpha$-MnTe, we know that $\gamma>0$,
giving the six degenerate states $\theta=\frac{\left(2n+1\right)\pi}{6}$,
with $n=0,\cdots,5$. The six domains are associated with the positive and negative signs of the magnetization $M_{z}$, corresponding to $n$ even and $n$ odd, and with the three possible orthorhombic domains from Eq. (\ref{eq:epsilon}), corresponding to the pairs $n=(0,3)$, $n=(4,1)$, and $n=(2,5)$. 

For simplicity, let us assume that the altermagnetic transition is approached from above $T_{\mathrm{AM}}$. Then, the dominant term in Eq. (\ref{eq:F_tot}) that sets the angle $\theta$ is the term with coefficient $\lambda_{1}$, as it is quadratic in $L$. This term favors $\theta=\alpha+m\pi$, for $\lambda_{1}\epsilon<0$, and $\theta=\alpha+\frac{\pi}{2}+m\pi$, for $\lambda_{1}\epsilon>0$, with $m=0,1$. The twofold degeneracy in $m$ reflects the fact that the two time-reversal related domains
are not affected by strain. For $\alpha=\pi/2$ (strain along $y$), this means that $\lambda_{1}\epsilon<0$ chooses two of the states that are also minimized by the sixth-order term (with coefficient $\gamma>0$), namely, $\theta=\pi/2$ and $\theta=3\pi/2$, corresponding to magnetic moments along $\pm y$. As a result, $\lambda_{1}\epsilon<0$ selects the single orthorhombic domain $n=(4,1)$.
For $\lambda_{1}\epsilon>0$, the states minimized by the dominant term are $\theta=0$ and $\theta=\pi$, which are maximally penalized by the sixth-order term. Because of this competition, the
system is expected to undergo another phase transition at a lower temperature, where $\theta$ moves to one of the four minima of the sixth-order term corresponding to $n=0,2,3,5$. Hence, $\lambda_{1}\epsilon>0$ selects the two orthorhombic domains $n=(0,3)$ and $n=(2,5)$ or, equivalently, disfavors the orthorhombic domain $n=(4,1)$. We note that there is an additional term in Eq. (\ref{eq:F_tot}) that depends linearly on strain, which is the term with coefficient $\lambda_{2}$. Generally, this term is sub-leading with respect to the term with coefficient $\lambda_{1}$, as it has an additional $L^{2}$ dependence.

In our neutron scattering experiments, compressive strain ($\epsilon<0$) along the $y$-axis (NNN direction) results in a sample with two orthorhombic domains. We therefore conclude that $\lambda_{1}\epsilon > 0$, implying that $\lambda_1 < 0$. The fact that a single domain is not observed upon applying tensile strain along the $y$-axis is likely a consequence of the smaller values of tensile strain that can be achieved in our setup, combined with the existence of a spinodal line for the two-domain structure. However, a detwinned single-domain is observed upon application of compressive strain along the $x$-axis (NN direction). This is consistent with our analysis, since due to the Poisson ratio, compressive strain along $x$  induces tensile strain along $y$.

The combination of strain and SOC also generates a new term in the AHE. We find that Eq. (\ref{eq:Mz}) is changed to:
\begin{equation}
\sigma_{xy}= A L^{3}\sin3\theta+B L\epsilon\sin\left(\theta+2\alpha\right)
\end{equation}
where $A$  and $B$ are temperature-dependent parameters. Let us analyze what happens to this equation when strain is applied along the $y$-axis, such that $\alpha=\pi/2$. As discussed above, there are in principle three orthorhombic domains corresponding to  $n=(0,3)$, $n=(4,1)$, and $n=(2,5)$, where $n$ labels the free energy minimum $\theta_n=\frac{\left(2n+1\right)\pi}{6}$ corresponding to magnetic moments along the $y$ direction. The applied magnetic field selects a single magnetic domain, corresponding to either $n$ even or $n$ odd. For concreteness, let us analyze the magnetic domain with $n$ even, such that the three orthorhombic domains become $n=0,2,4$. Averaging over the domains gives:
\begin{equation}
\label{eq:sigmaxy}
\begin{aligned}
\bar{\sigma}_{xy} &= A L^{3} \left[\frac{1}{3} \sum_{n=0,2,4}(-1)^n \right]\\
&- B L\epsilon \left[\frac{1}{3} \sum_{n=0,2,4}\sin \frac{(2n+1)\pi}{6} \right]
\end{aligned}
\end{equation}
The last term vanishes, implying that the linear-in-$L$ contribution to the domain-averaged anomalous Hall conductivity vanishes (the cubic-in-$L$ contribution is unaffected by the domain average). However, in our transport experiments, a single domain is achieved either by applying compressive strain along the NN direction or due to the built-in tensile strain along the NNN direction from the thermal coefficient mismatch between the sample and the strain cell. As a result, the expression above is modified since only the $n=4$ domain contributes, yielding:

\begin{equation}
\sigma_{xy} = A L^{3} +B L\epsilon 
\end{equation}

This linear relationship between $\sigma_{xy}$ and $\epsilon$ is the same as the one observed experimentally.  Note that this expression remains valid when $\epsilon$  changes sign as long as the single domain state survives, i.e., if the strain value is such that the corresponding spinodal line is not crossed.  The main consequence of this expression is that, near the temperature regime where the temperature-dependent parameter $A$ changes, the contribution for the first term is small  and can thus be overcome by the second contribution, leading to a sign change of the AHE for the appropriate sign of $\epsilon$.

\section*{Data Availability}
All data that support the findings of this paper are included in the main text and Methods. Source data for transport measurements are provided with this paper. Neutron scattering source data are available upon request from the corresponding authors.

\section{Competing interests}
The authors declare no competing interests.

\section*{Acknowledgments} 
We thank K. D. Belashchenko,  Hua Chen, and M. Khodas for helpful discussions. 
The single-crystal synthesis, transport, and neutron scattering experiments at Rice were supported by the U.S. DOE, BES under Grant Nos. DE-SC0012311 (P.D.) and DE-SC0026179 (P.D., E.M.). Part of the materials characterization efforts at Rice is supported by the Robert A. Welch Foundation Grant No. C-1839 (P.D.). The elastocaloric effect and transport measurements at the University of Washington is supported by the Air Force Office of Scientific Research under grant FA9550-21-1-0068, the David and Lucile Packard Foundation and partially supported by NSF through the University of Washington Molecular Engineering Materials Center, a Materials Research Science and Engineering Center (DMR-2308979). 
Work at UH is supported by the Enterprise Science Fund of Intellectual Ventures Management, LLC; U.S. Air Force Office of Scientific Research Grants FA9550-15-1-0236 and FA9550-20-1-0068; the T. L. L. Temple Foundation; the John J and Rebecca Moores Endowment; the Robert A. Welch Foundation (00730-5021-H0452-B0001-G0512489); and the State of Texas through the Texas Center for Superconductivity at the University of Houston. R.M.F. (theoretical model) was supported by the Air Force Office of Scientific Research under Award No. FA9550-21-1-0423.
A portion of this work was performed at the National High Magnetic Field Laboratory, which is supported by the National Science Foundation Cooperative Agreement No. DMR-2128556 and the State of Florida.
A portion of this research used resources at the Spallation Neutron Source, a DOE Office of Science User Facility operated by the Oak Ridge National Laboratory. The beam times were allocated to CORELLI, VERITAS (HB1A), and SEQUOIA on Proposals No. IPTS-36272, IPTS-34643, and IPTS-34120, respectively.

\section*{Contributions}
Z.L. and P.D. conceived the project. S.X. and Z.L. synthesized the crystals and carried out the measurements, with assistance from J.D., E.R., J.L., R.C., Y.Z., S.P., C.-W.C., and L.Z.D. 
Li-doped MnTe samples for initial neutron scattering measurements were provided by E.M. group. 
Neutron scattering experiments under uniaxial stress were designed by S.X and Z.L., and carried out by T.Z., F.Y., W.T., M.B.S., and S.X.. J.D., E.R., and S.E. performed the elastocaloric effect measurements under the supervision of J.-H.C. The data were analyzed by S.X and Z.L.. R.M.F. developed the theoretical model.
Z.L., J.-H.C., and P.D. supervised the work. The paper is written by P.D., Z.L., S.X., R.M.F., and J.-H.C. with input from all coauthors.


\bibliography{MnTe_Ref_1}

\clearpage
\pagebreak
\setcounter{page}{1}

\renewcommand{\figurename}{\textbf{Extended Data} Fig.}
\renewcommand\thefigure{\arabic{figure}}
\setcounter{figure}{0}
\setcounter{equation}{0}

\begin{figure*}
    \centering
    \includegraphics[width = 183mm]{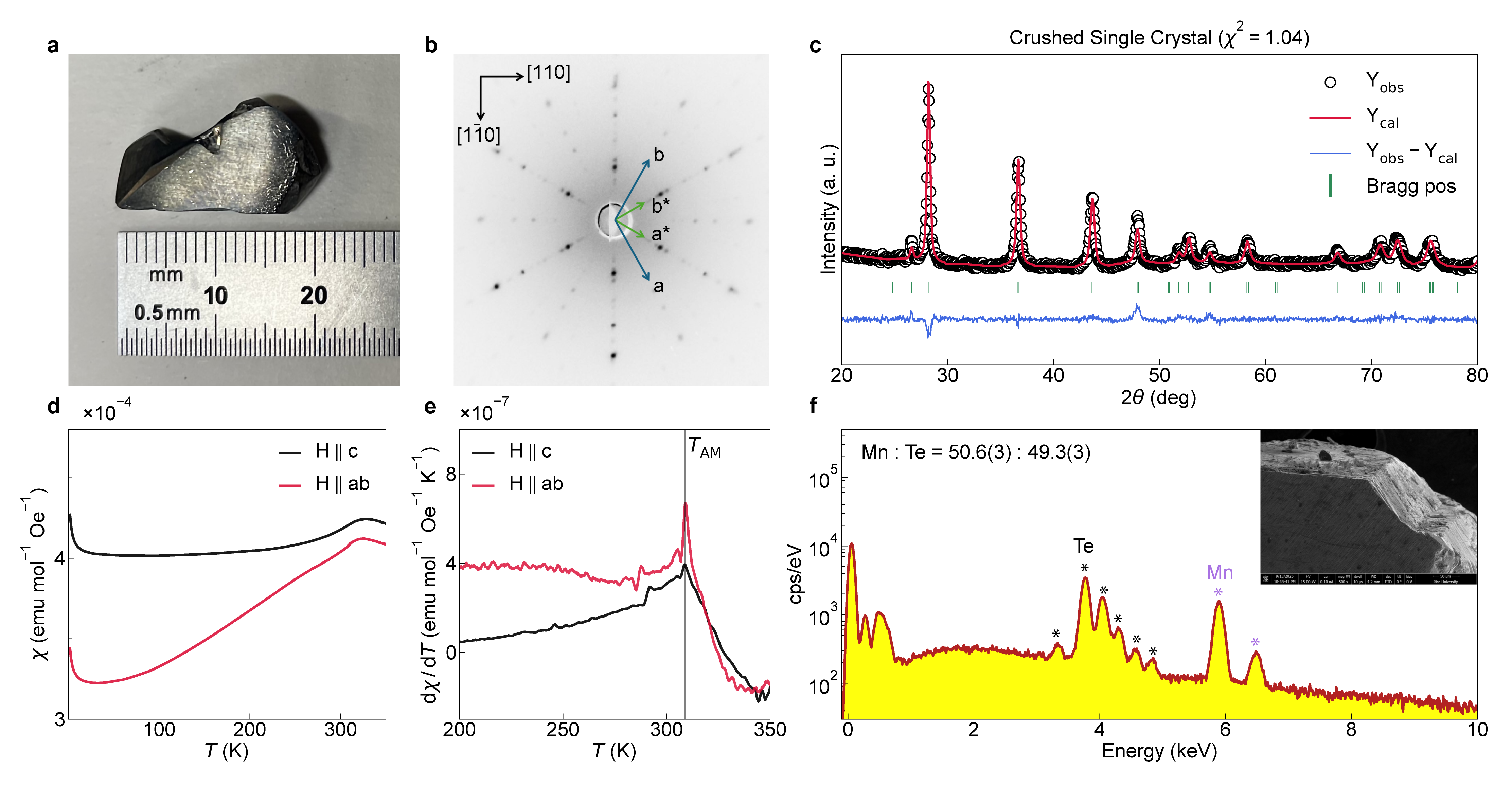}    
    \caption{\textbf{Characterizations of single crystal hexagonal MnTe.}  
    \textbf{a,} Single crystal image.
    \textbf{b,} X-ray Laue diffraction pattern of a $\alpha$-MnTe single crystal. The crystallographic orientations along the $[1,1,0]$ and $[1,\bar{1},0]$ directions in real space are marked. Real-space axes ($\textbf{a}$, $\textbf{b}$) and corresponding reciprocal lattice vectors ($\textbf{a}^*$, $\textbf{b}^*$) are marked by blue and green arrows, respectively.
    \textbf{c,} Powder X-ray diffraction pattern at room temperature. Black circles represent experimental intensities and the red line shows the calculated intensities. The refinement yields $\chi^2 = 1.04$, confirming the pure $\alpha$-MnTe phase.
    \textbf{d,} Temperature dependence of magnetic susceptibility with magnetic fields applied along the in-plane and out-of-plane directions.
    \textbf{e,} Derivative of data in \textbf{d}, where the peak features mark the altermagnetic (AM) transition $T_{\mathrm{AM}}$.
    \textbf{f,} Energy-dispersive X-ray spectroscopy at room temperature. Tellurium and manganese peaks are marked by black and purple stars, respectively. The measured Mn:Te ratio is $1.026\pm0.014$.
    \label{figS:characterization}}
\end{figure*}

\begin{figure*}
    \centering
    \includegraphics[width = 150mm]{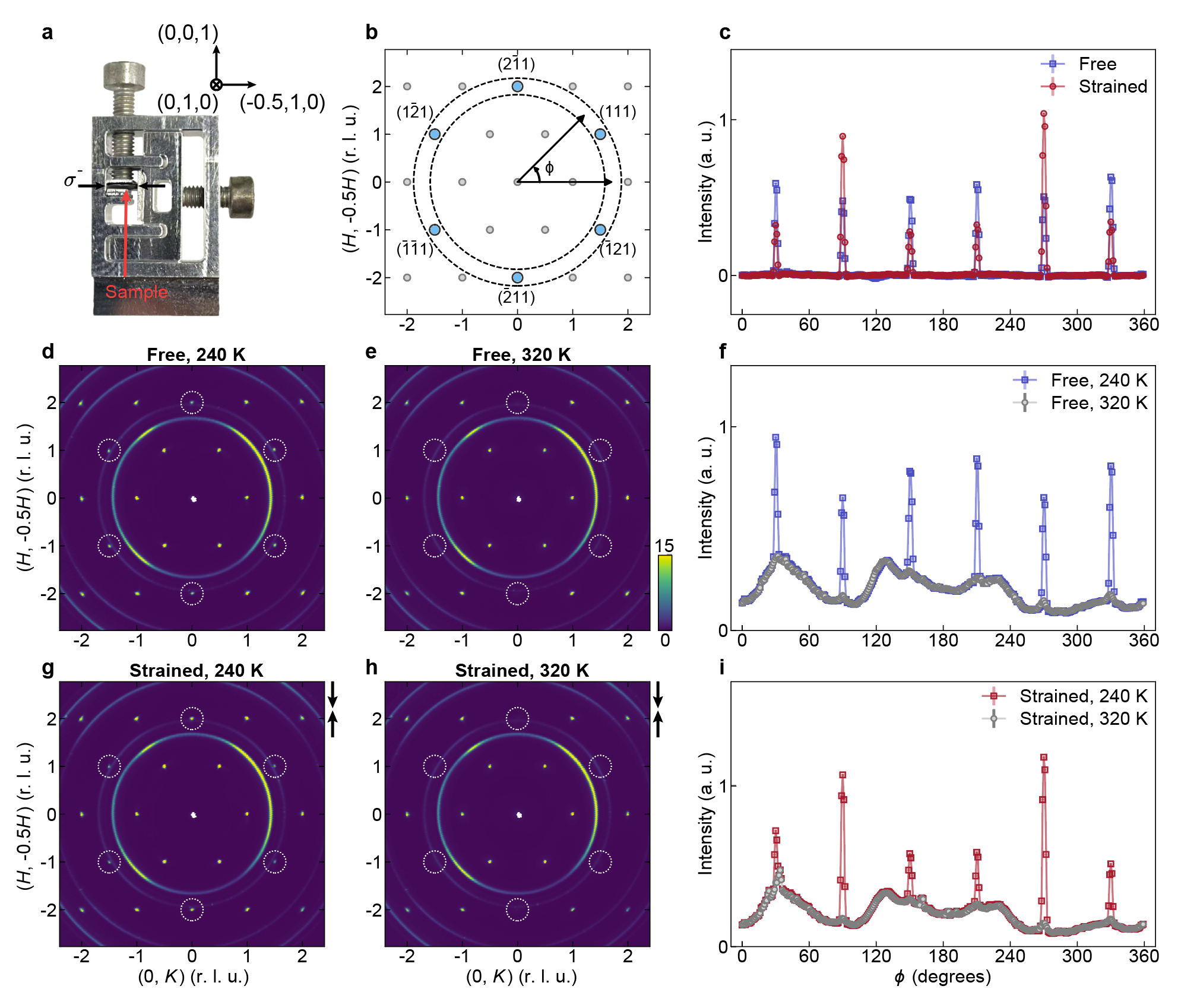} 
    
    \caption{\textbf{Neutron scattering signatures of strain applied along the nearest-neighbor (NN) Mn–-Mn direction in $\alpha$-MnTe.}
    \textbf{a,} Mechanical uniaxial stress device for neutron scattering experiments. A square plate-shaped sample with mass of $\sim$ 90 mg was mounted in the device as indicated by the red arrow. Compressive stress $\sigma^-$ was applied by tightening the screw on the right. The uniaxial stress is along the NN Mn--Mn bond direction in real space.
    \textbf{b,} Schematic of the $(H, -0.5H)\times(0,K)$ reciprocal space plane at $L=1$ (r.l.u.) in the altermagnetically ordered phase. Blue markers indicate the magnetic-only Bragg reflections used in our analysis, where nuclear contributions are absent, while gray markers denote Bragg positions with nuclear intensity.
    \textbf{c,} Azimuthal angular dependence of magnetic peak intensities of free-standing and NN-strained samples with background subtracted from the measurements at $320$ K ($>T_{\mathrm{AM}}$) . 
    \textbf{d-f,} Neutron diffraction patterns of the free-standing sample. The intensities were integrated within the annular region enclosed by the two dashed circles, restricted to the areas of the six magnetic Bragg peaks. The integration region is enlarged for clarity in \textbf{b}. 
    \textbf{g-i,} Similar plots to \textbf{d-f} for the same sample under compressive strain. The white arrows represent the uniaxial compressive strain direction.
    \label{figS:neutron_NN}}
\end{figure*}

\begin{figure*}
    \centering
    \includegraphics[width = 140mm]{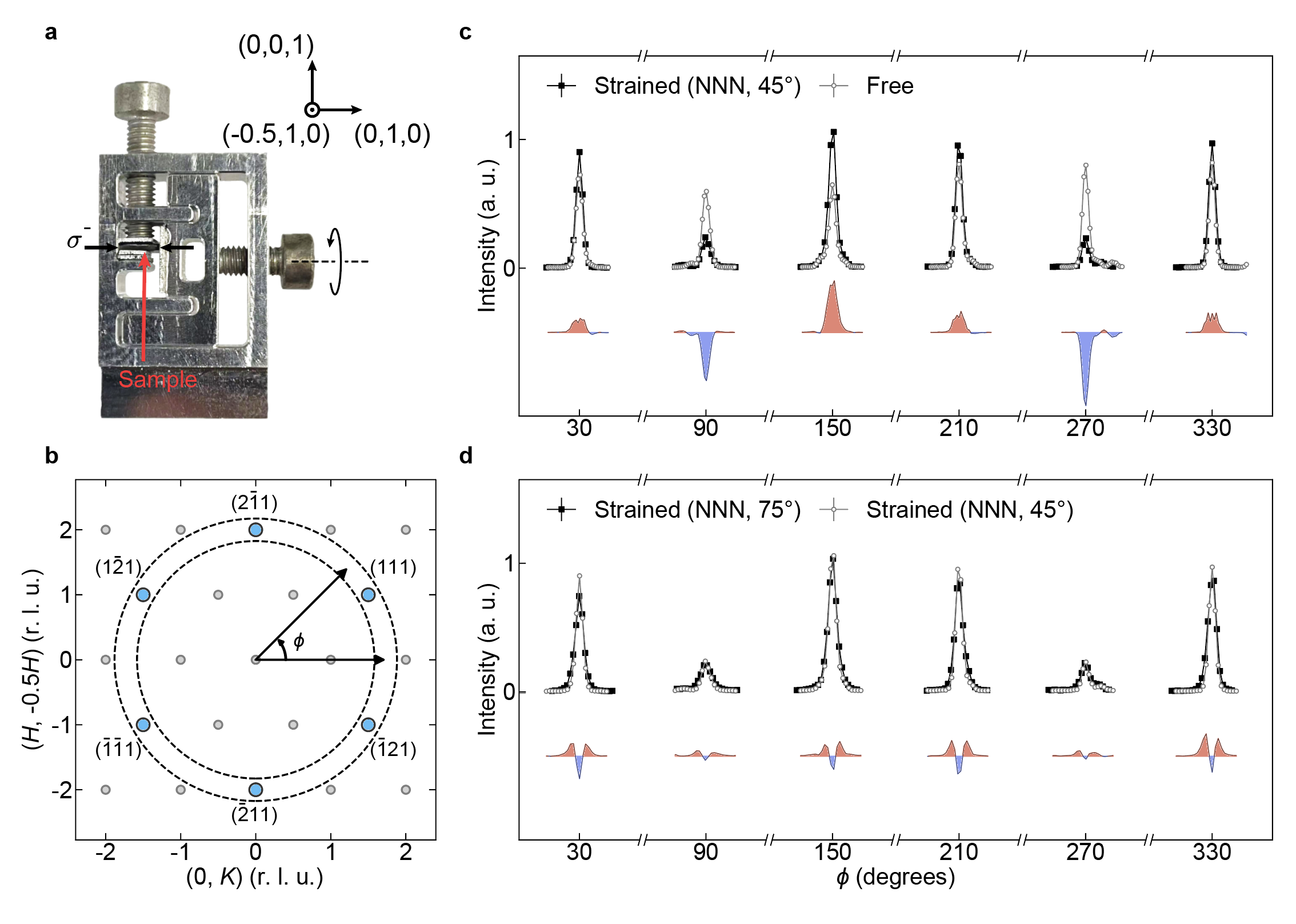} 
    \caption{
    \textbf{Neutron scattering signatures of strain applied along the next-nearest-neighbor (NNN) Mn-–Mn direction in $\alpha$-MnTe.}
    \textbf{a,} Mechanical uniaxial-stress device for neutron scattering experiments. A square, plate-shaped $\sim$ 75 mg sample is mounted as indicated by the red arrow. Compressive stress $\sigma^-$ is applied by tightening the screw on the right, loading the sample along the NNN Mn–Mn bond direction in real space.
    \textbf{b,} Schematic of the $(H, -0.5H)\times(0,K)$ reciprocal space plane at $L=1$ (r.l.u.) in the altermagnetically ordered phase. Blue markers indicate the magnetic-only Bragg reflections used in our analysis, where nuclear contributions are absent, while gray markers denote Bragg positions with nuclear intensity.
    \textbf{c,} Azimuthal dependence of magnetic peak intensities for the free-standing and NNN–strained samples (45$^\circ$ screw rotation), together with their difference. Color shading highlights that the two peaks perpendicular to the uniaxial-stress axis are suppressed (blue), while the remaining four are enhanced (red).
    \textbf{d,} Strain-dependence of the azimuthal profile. Increasing the applied strain (75$^\circ$ screw rotation) does not further modify the intensity ratios.
    \label{figS:neutron_NNN}}
\end{figure*}

\begin{figure*}
    \centering
    \includegraphics[width = 170mm]{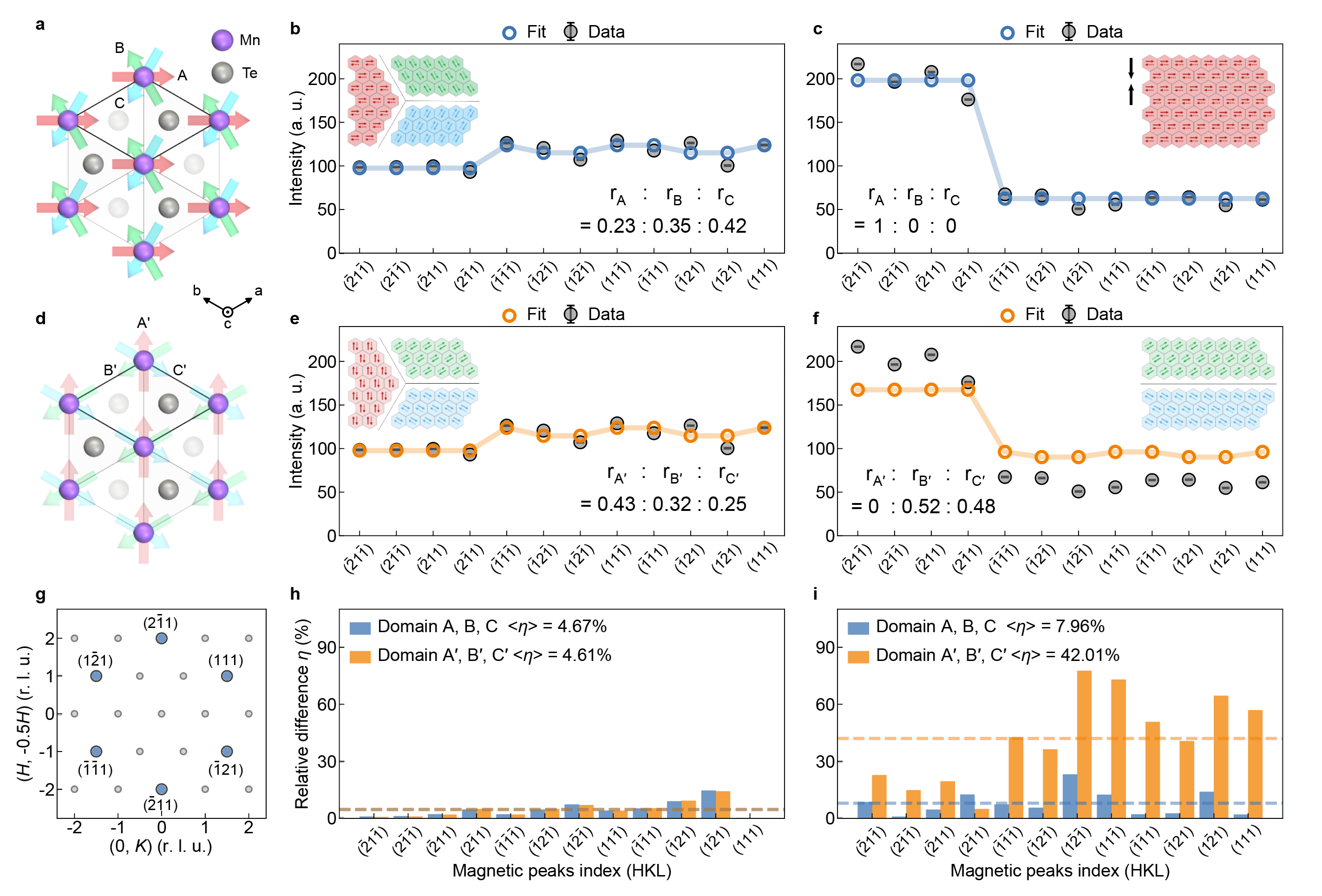} 
    \caption{\textbf{In-plane magnetic moment orientation determination in free-standing and strained $\alpha$-MnTe}  
    \textbf{a,} Magnetic structure of $\alpha$-MnTe with three in-plane magnetic domains $(\mathrm{A},\mathrm{B},\mathrm{C})$. The moments point along the crystallographic $[1,\bar{1},0]$, $[1,2,0]$, and $[-2,−1,0]$ directions, respectively. 
    \textbf{b,} Refinement of magnetic structure factors for selected Bragg reflections in free-standing $\alpha$-MnTe where the nuclear intensity is absent and the observed intensity is purely magnetic, yielding the domain population ratios for $\mathrm{A}$, $\mathrm{B}$, and $\mathrm{C}$.
    \textbf{c,} Corresponding refinement for strained $\alpha$-MnTe, showing a predominant population of domain $\mathrm{A}$.
    \textbf{d,} Magnetic structure of $\alpha$-MnTe with three in-plane magnetic domains $(\mathrm{A}^\prime,\mathrm{B}^\prime,\mathrm{C}^\prime)$, where the moments point along the crystallographic $[1,1,0]$, $[\bar{1},0,0]$, and $[0,\bar{1},0]$ directions, respectively.
    \textbf{e,f,} Refinement on the same dataset as in \textbf{b,c} assuming domain $\mathrm{A}^{\prime}$, $\mathrm{B}^{\prime}$, and $\mathrm{C}^{\prime}$. 
    \textbf{g,} Schematic of magnetic Bragg reflections with no nuclear contribution (blue dots) in the $L=1$ (r.l.u.) plane. The corresponding $(H,K)$ reflections at $L=-1$ (r.l.u.) are also included in the refinement.
    \textbf{h}, Relative difference $\eta = \lvert I_{\mathrm{Data}}-I_{\mathrm{Fit}}\rvert / I_{\mathrm{Data}} \times 100\,\%$ for each reflection comparing the two domain models shown in \textbf{a} and \textbf{d} in the free-standing $\alpha$-MnTe. The dotted lines indicate the average relative differences among all 12 peaks using the two models. Both models describe the data comparably well in the free-standing case.
    \textbf{i}, Corresponding comparison in the strained case, where the NNN Mn--Mn bond moment direction domain model $(\mathrm{A},\mathrm{B},\mathrm{C})$ shown in \textbf{a} provides the better fit.
    \label{figS:simulation}}
\end{figure*}

\begin{figure*}
    \centering
    \includegraphics[width = 110mm]{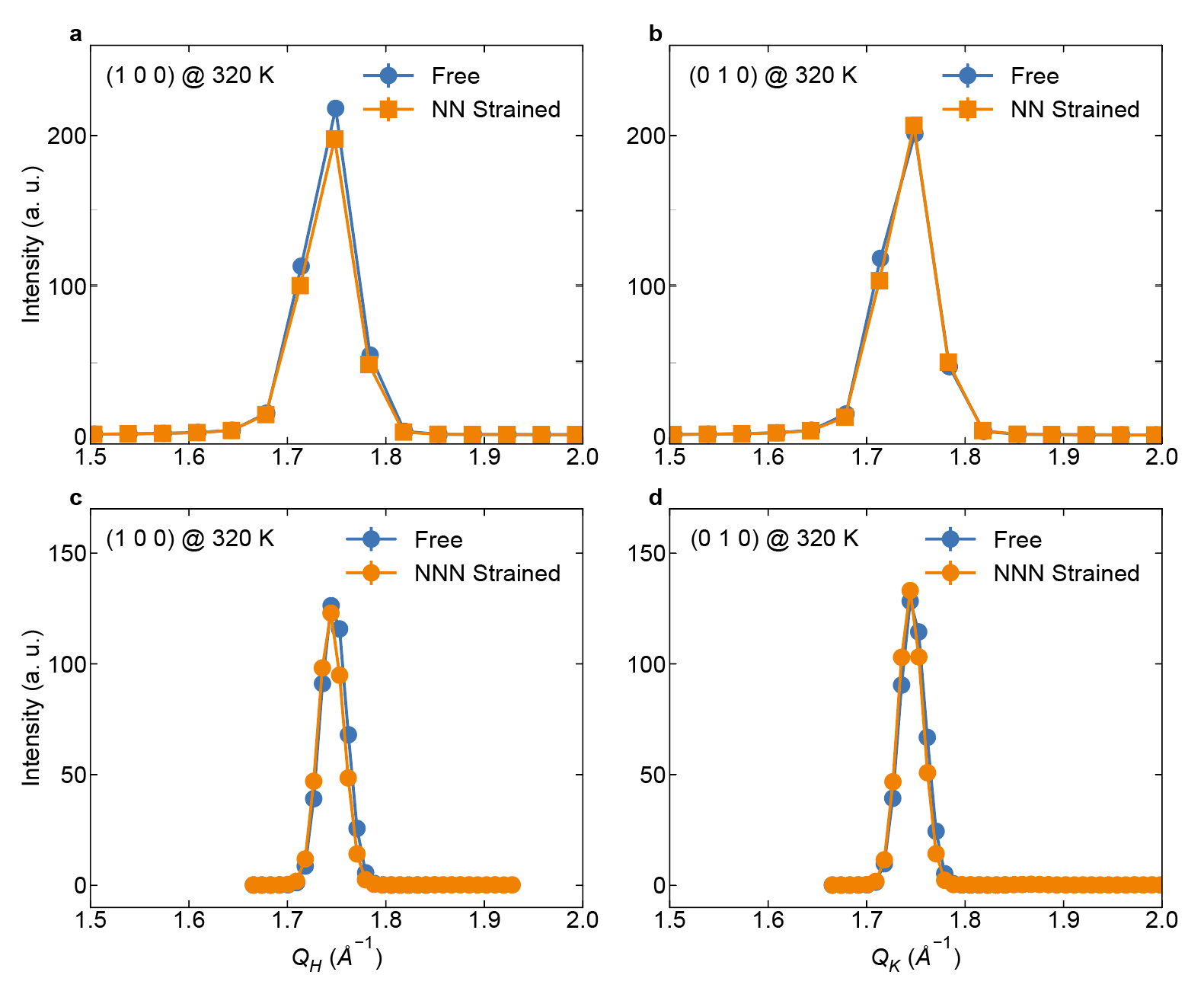}
    \caption{\textbf{Nuclear Bragg peak variations under strain.}
    \textbf{a, b,} NN-strain dependence of the nuclear Bragg peak intensities at 
    $\mathbf{Q}=(100)$ (\textbf{a}) and $\mathbf{Q}=(010)$ (\textbf{b}) measured at 320 K on CORELLI. 
    \textbf{c, d,} NNN-strain dependence of the nuclear Bragg peak intensities at 
    $\mathbf{Q}=(100)$ (\textbf{c}) and $\mathbf{Q}=(010)$ (\textbf{d}) measured at 320 K on HB1A. 
    Within our accessible strain range, no lattice-parameter change was resolved in either strain direction, implying that any distortion is below the typical neutron diffraction resolution.
}
    \label{figS:Braggstrain}
\end{figure*}

\begin{figure*}
    \centering
    \includegraphics[width = 183mm]{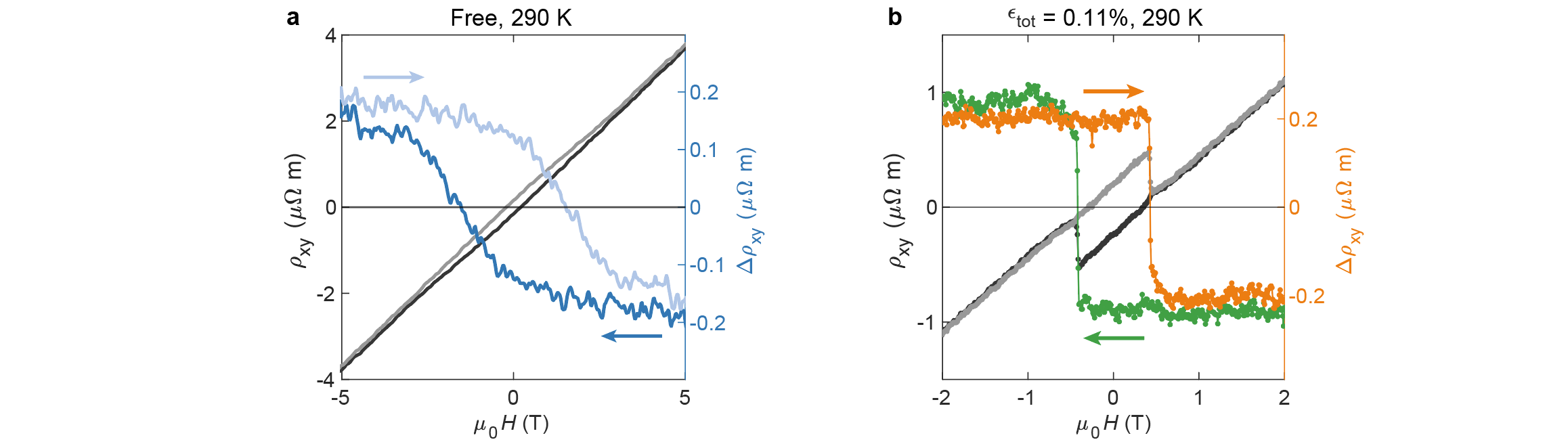}    
    \caption{\textbf{Linear subtraction of the AHE in free-standing $\alpha$-MnTe.}  
    \textbf{a,} Magnetic field dependence of the Hall resistivity $\rho_{xy}$ and $\Delta\rho_{xy}$ after a linear background subtraction at 290 K in the free-standing sample. 
    \textbf{b,} Similar plot to \textbf{a}, obtained from the same sample under uniaxial strain of $\epsilon_{\mathrm{tot}} = 0.11\,\%$ along the NNN Mn--Mn bond direction.
    \label{figS:AHEsubtraction}}
\end{figure*}

\begin{figure*}
    \centering
    \includegraphics[width = 120mm]{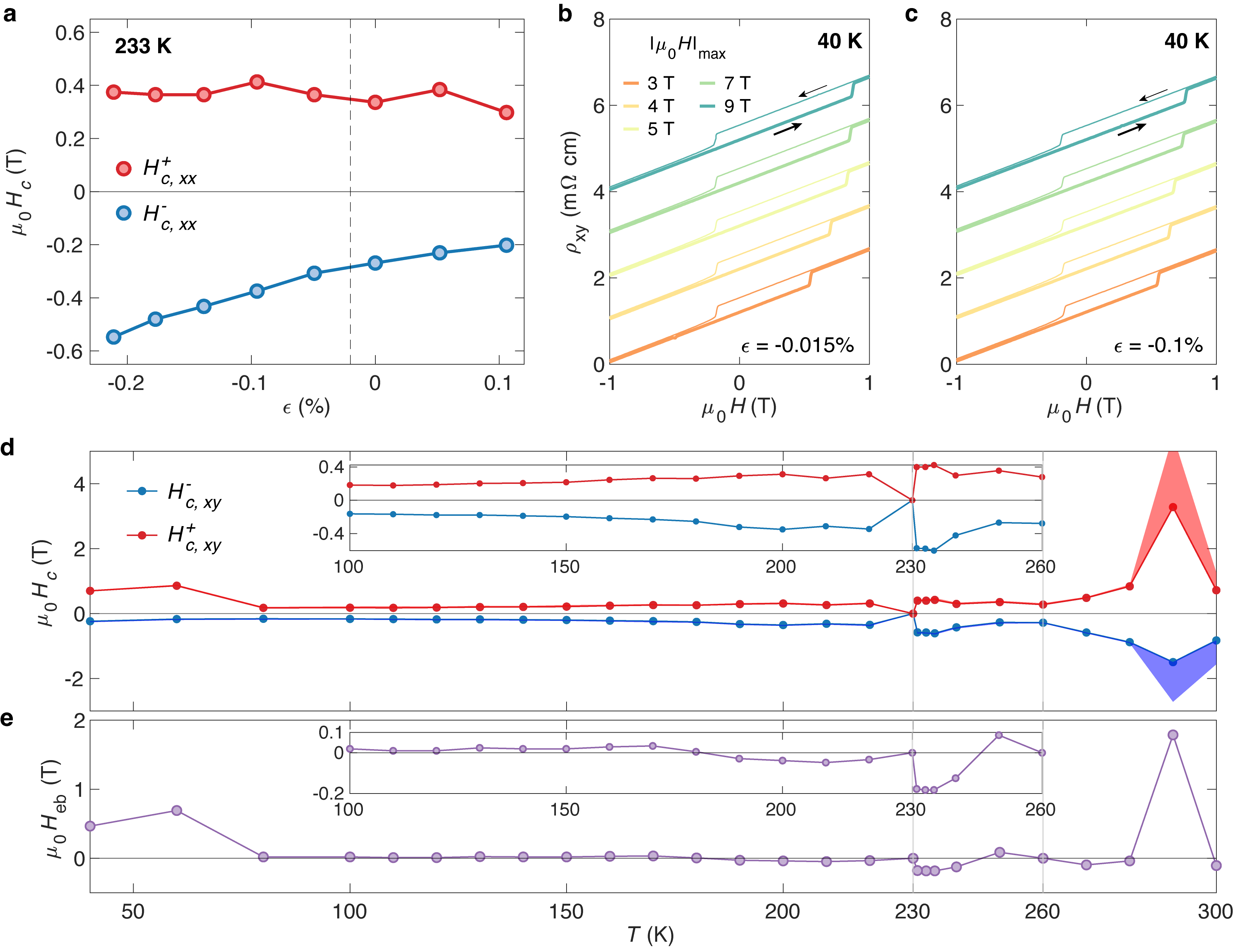}    
    \caption{\textbf{Coercive field as function of strain and temperature.}  
    \textbf{a,} Strain dependence of the coercive fields extracted from $\rho_{xx}$ at 233 K. Both $H_c^{\pm}$ evolve smoothly with strain across the AHE vanishing point, marked by the dashed line.
    \textbf{b,c,} Hall resistivity as a function of magnetic field at 40 K under strains of $\epsilon=-0.015\,\%$ and $\epsilon=-0.1\,\%$, measured with different maximum field ranges. The exchange bias effect is enhanced by increasing $\mu_0\it H_{\mathrm max}$ and applying compressive uniaxial strain. 
    \textbf{d,e,} Temperature dependence of the coercive fields $H_{c,xy}^-$ and $H_{c,xy}^+$ and the exchange bias $H_{eb} = H_{c,xy}^- + H_{c,xy}^+$ at $\epsilon_{\rm Max}^-$, measured under a field sweep range of $\pm 9$ T. The exchange bias becomes increasingly pronounced at low temperatures, likely due to stronger magnetic coupling. Inset focus on the range from 100 K to 260 K.
    \label{figS:exchangebias}}
\end{figure*}

\begin{figure*}
    \centering
    \includegraphics[width = 120mm]{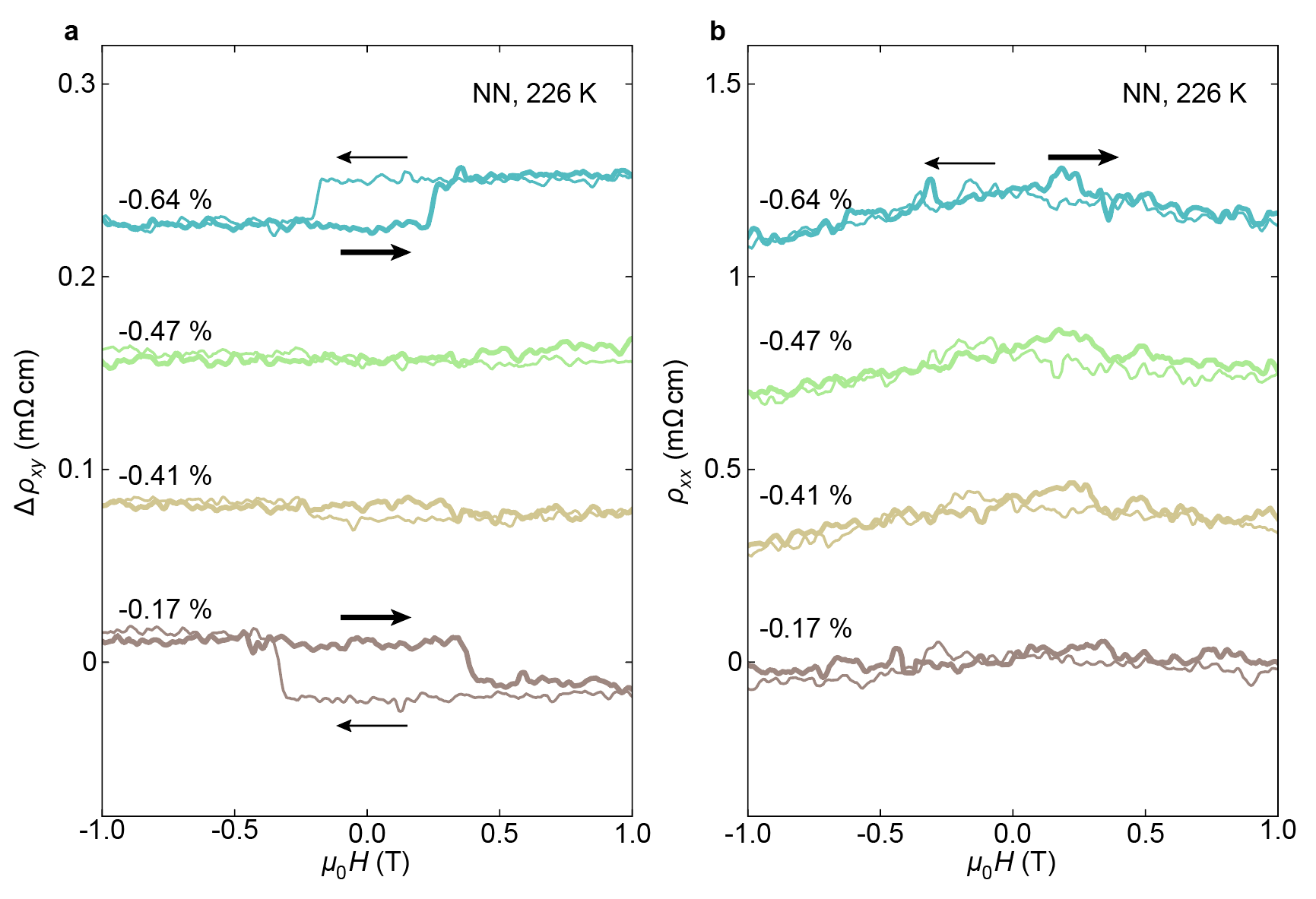}    
    \caption{\textbf{Sign reversal of the AHE in $\alpha$-MnTe observed in NN configuration.}
    \textbf{a,} $\Delta\rho_{xy}$ (linear background subtracted) at 226 K under different strain levels. The legend denotes the total strain experienced by the crystal. The signal shows a sign reversal similar to the NNN sample in Fig.~\ref{fig:fig3}d. Thin (thick) black arrows denote magnetic-field sweeps down (up).
    \textbf{b,} Corresponding longitudinal resistivity $\rho_{xx}$ measured under the same strain and temperature as in \textbf{a}. All the curves are vertically offset for clarity.
    \label{figS:nnsignchange}}
\end{figure*}

\begin{figure*}
    \centering
    \includegraphics[width = 150mm]{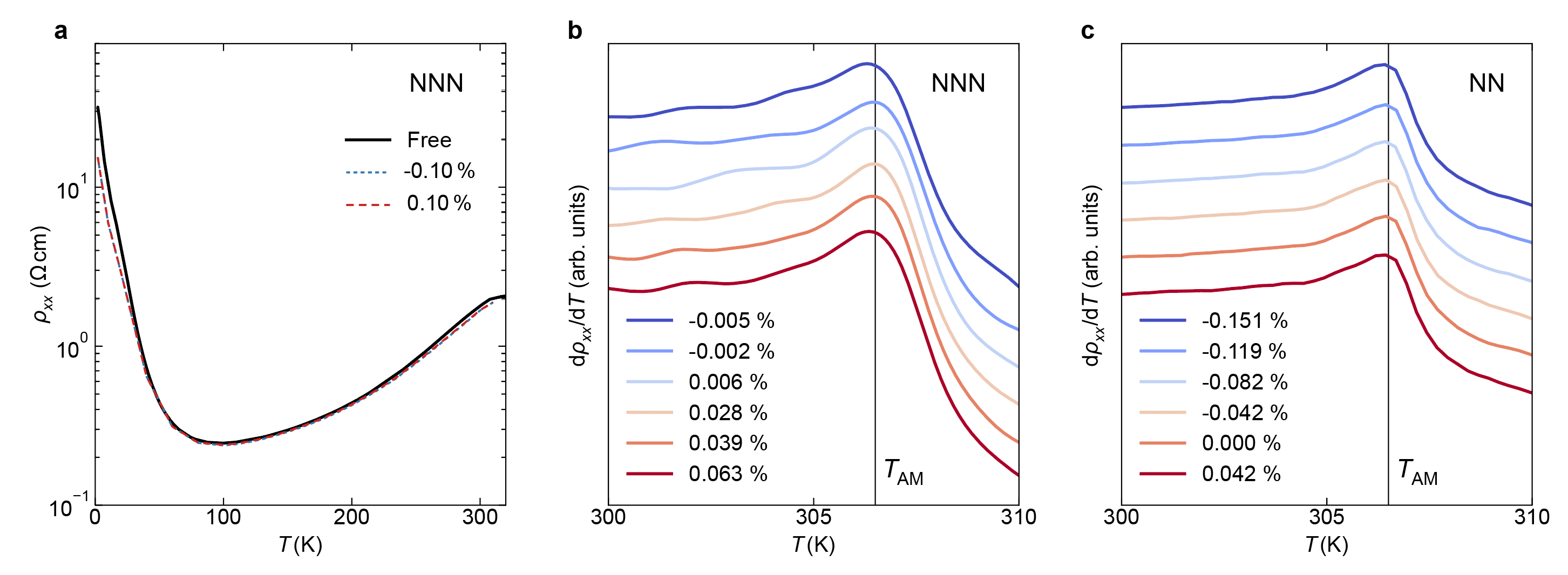}    
    \caption{\textbf{Strain-tuning longitudinal resistivity in $\alpha$-MnTe.}  
    \textbf{a,} Temperature dependence of longitudinal resistivity $\rho_{xx}$ for free-standing and strained samples. No obvious change in resistivity is observed under our experimental strain conditions, confirming that the AHE variations are not driven by longitudinal conductivity changes.
    \textbf{b,} First derivative of resistivity versus temperature at different strain levels. The peak corresponding to the altermagnetic transition temperature $T_{\mathrm{AM}}$ remains unchanged as the uniaxial strain is varied from compressive to tensile.
    \textbf{c,} Same plot as in \textbf{b}, in NN configuration.
    \label{figS:dRdT}}
\end{figure*}

\begin{figure*}
    \centering
    \includegraphics[width = 130mm]{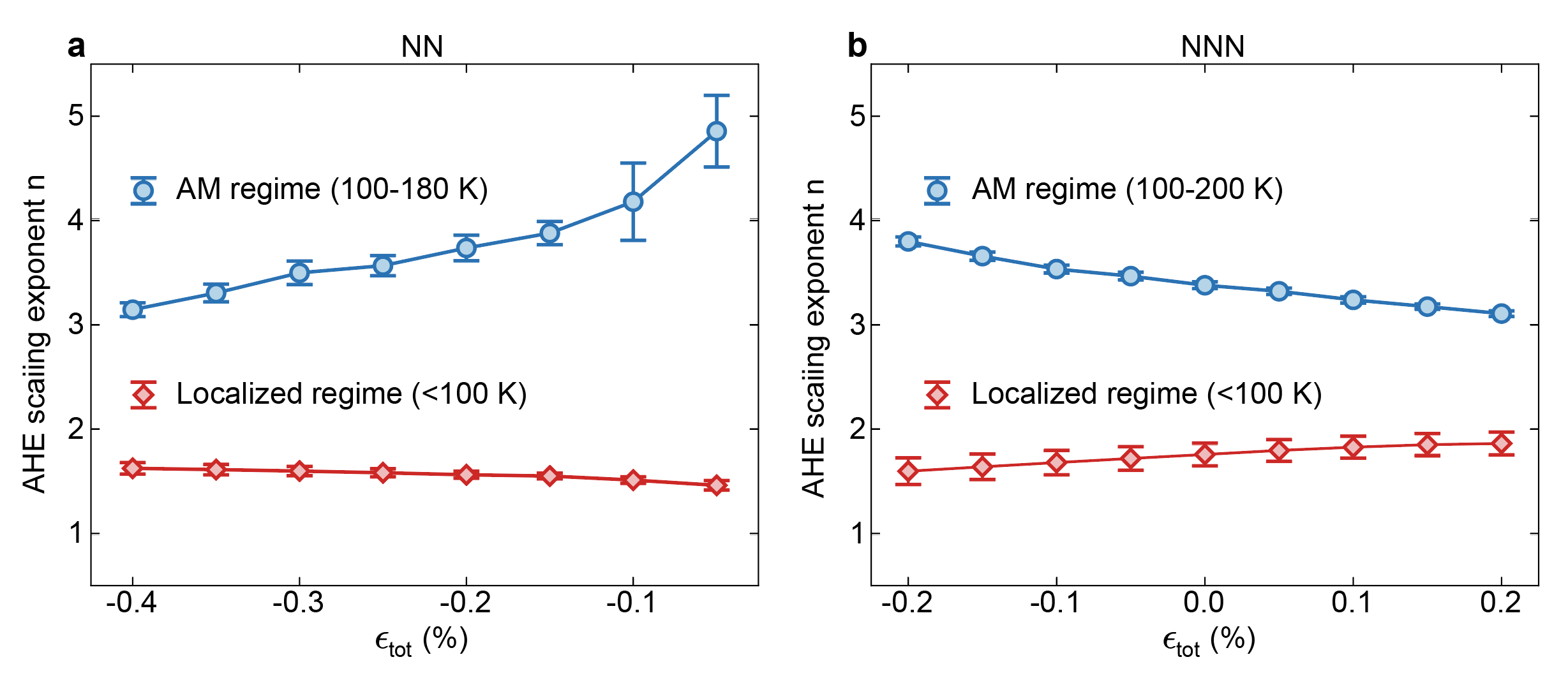}    
    \caption{\textbf{Strain dependence of the anomalous Hall conductivity scaling exponent $n$} in $|\sigma_{xy}^{\mathrm{A}}|\propto\sigma_{xx}^n$ for two samples measured in the NN (\textbf{a}) and NNN (\textbf{b}) configurations. In the localized hopping regime ($T<100$ K), $n$ is nearly strain independent and remains in the range $\sim$1.5–1.8 for both samples. In the AM regime ($100<T<180$ K), $n$ decreases and then saturates toward $n\approx 3$ as the strain is tuned towards the direction that stabilizes the single in-plane domain state: for the NN sample, this occurs on the compressive-NN side, while for the NNN sample it occurs on the tensile-NNN side (equivalently, compressive-NN). Notably, $n$ continues to evolve with strain even after detwinning, indicating strain tuning within a single in-plane domain.
    }
    \label{figS:AHCscaling}
\end{figure*}

\begin{figure*}
    \centering
    \includegraphics[width = 180mm]{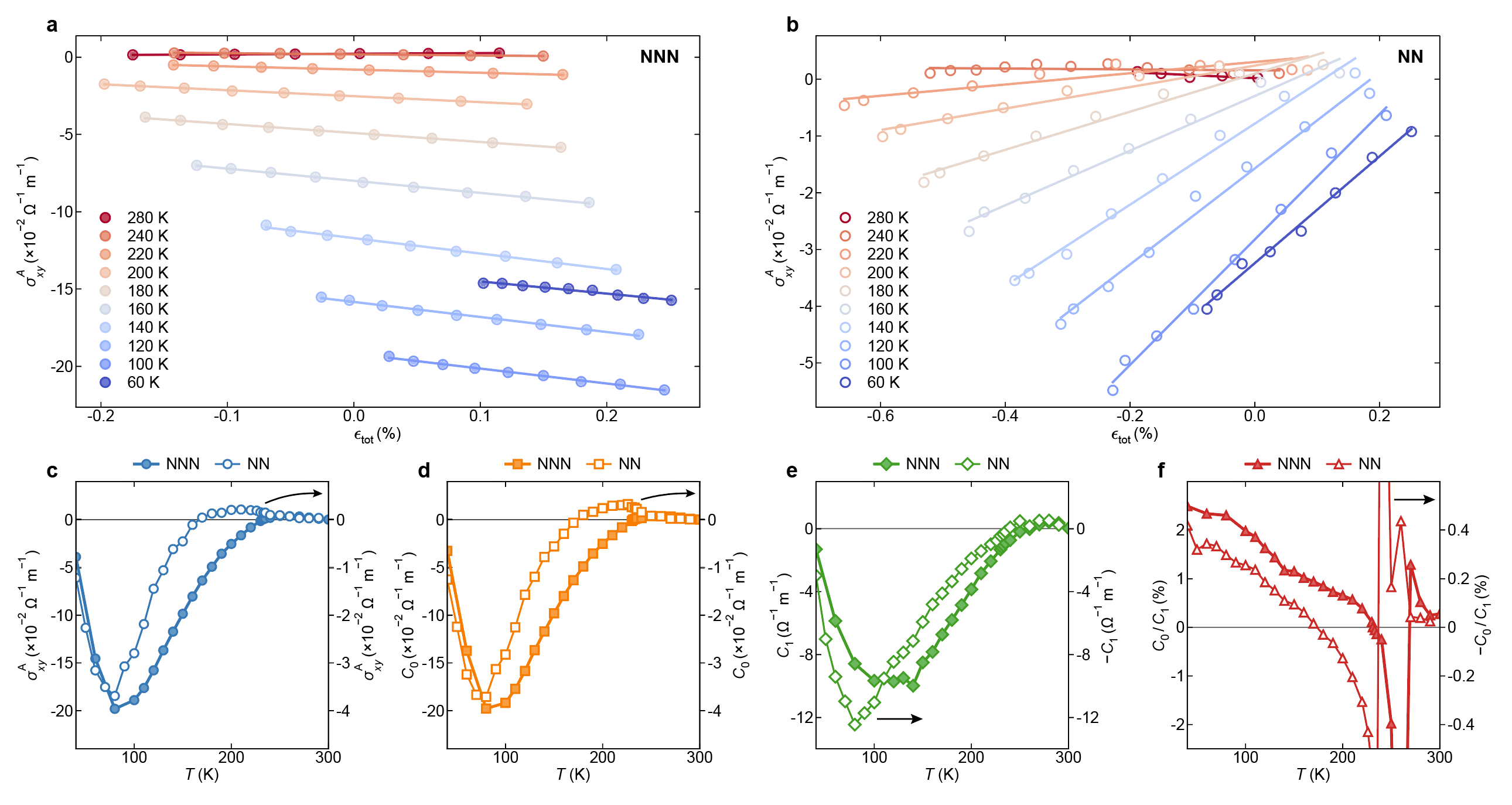}  
    \caption{\textbf{Separation of the strain-induced contributions to the AHC.}
    \textbf{a--b,} Anomalous Hall conductivity $\sigma_{xy}^{\mathrm A}$ as a function of strain applied along NNN and NN directions at different temperatures. The solid lines are linear fits using $\sigma_{xy}^{A}(\epsilon,T)=C_0(T)+C_1(T)\,\epsilon$. Within our phenomenological model, $C_0$ represents the intrinsic contribution driven by spin--orbit coupling ($C_0=A\,L^3$), whereas $C_1$ represents the strain-induced contribution ($C_1 =  B L$). In each panel, $\epsilon>0$ refers to tensile strain along the corresponding direction. Note that $\epsilon$ is the total strain experienced by the sample, which includes the temperature-dependent built-in tensile strain induced by the thermal expansion mismatch, as shown in Fig.\ref{figS:Silicon}.
    \textbf{c,} Temperature dependence of the zero applied strain anomalous Hall conductivity (AHC).
    \textbf{d--e,} Temperature dependence of the coefficients $C_0$ and $C_1$ obtained from the linear fits in panels \textbf{a} and \textbf{b}. Note the opposite signs of $C_1$ for strain applied along the NN and NNN directions.
    \textbf{f,} Temperature dependence of the ratio $C_0/C_1$.}
    \label{figS:model_fit}
\end{figure*}

\begin{figure*}
    \centering
    \includegraphics[width = 130mm]{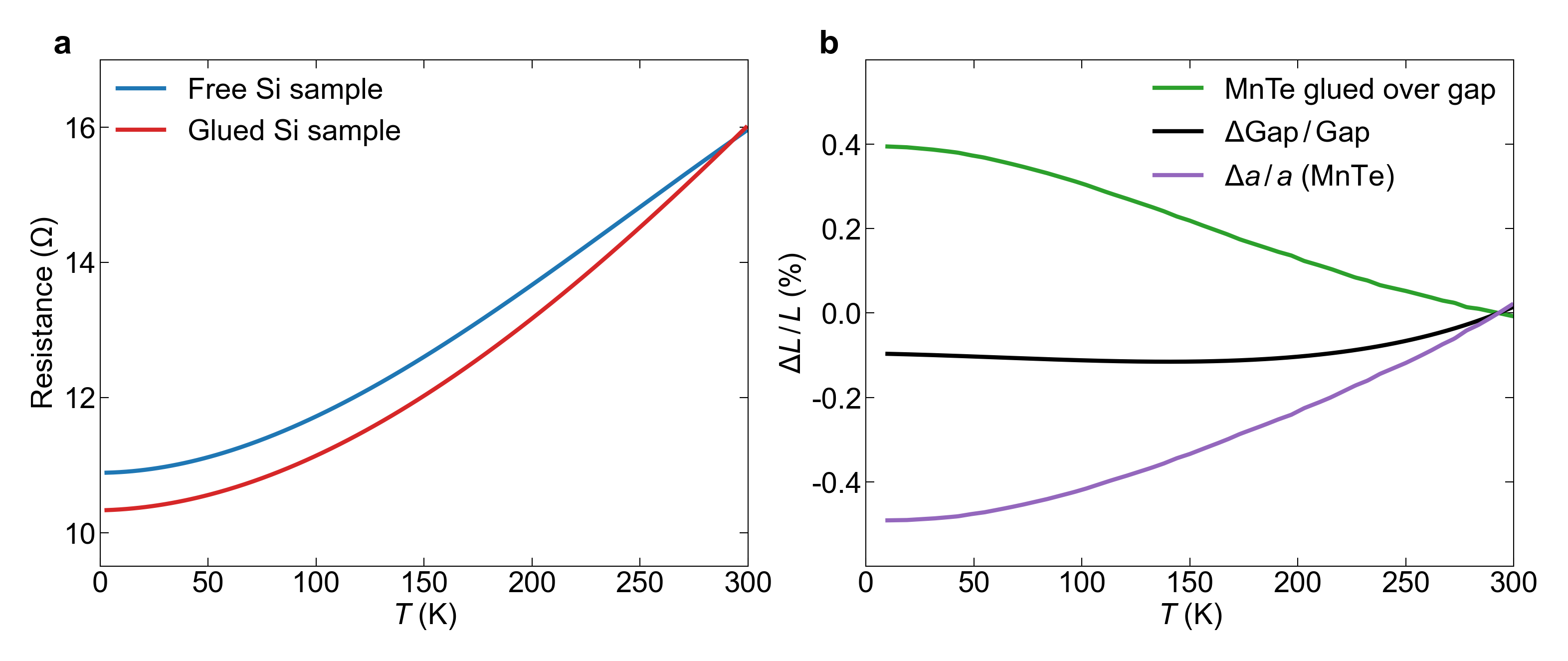}
    \caption{\textbf{Estimate of thermally induced strain using a silicon gauge.}
    \textbf{a,} Silicon resistance as a function of temperature measured in a freestanding configuration (one end glued; blue) and in the fully glued configuration (both ends glued; red). The relative resistance change with respect to 293\,K, $\Delta R/R_{293\mathrm{K}}$, is converted to a relative gap change using the silicon gauge factor $g$ (Methods).
    \textbf{b,} Relative length change versus temperature. The gap variation $\Delta\mathrm{Gap}/\mathrm{Gap}$ is obtained from \textbf{a}. The MnTe lattice contraction $\Delta a/a$ is calculated from Ref.~\cite{baral2023giant}. The thermally induced strain on a MnTe crystal bridging the gap is estimated as $\epsilon_{\mathrm{th}}=\Delta\mathrm{Gap}/\mathrm{Gap}-\Delta a/a$, yielding an approximate tensile strain of $\sim$ 0.2\,\% near 200\,K.}
    \label{figS:Silicon}
\end{figure*}

\begin{figure*}
    \centering
    \includegraphics[width = 140mm]{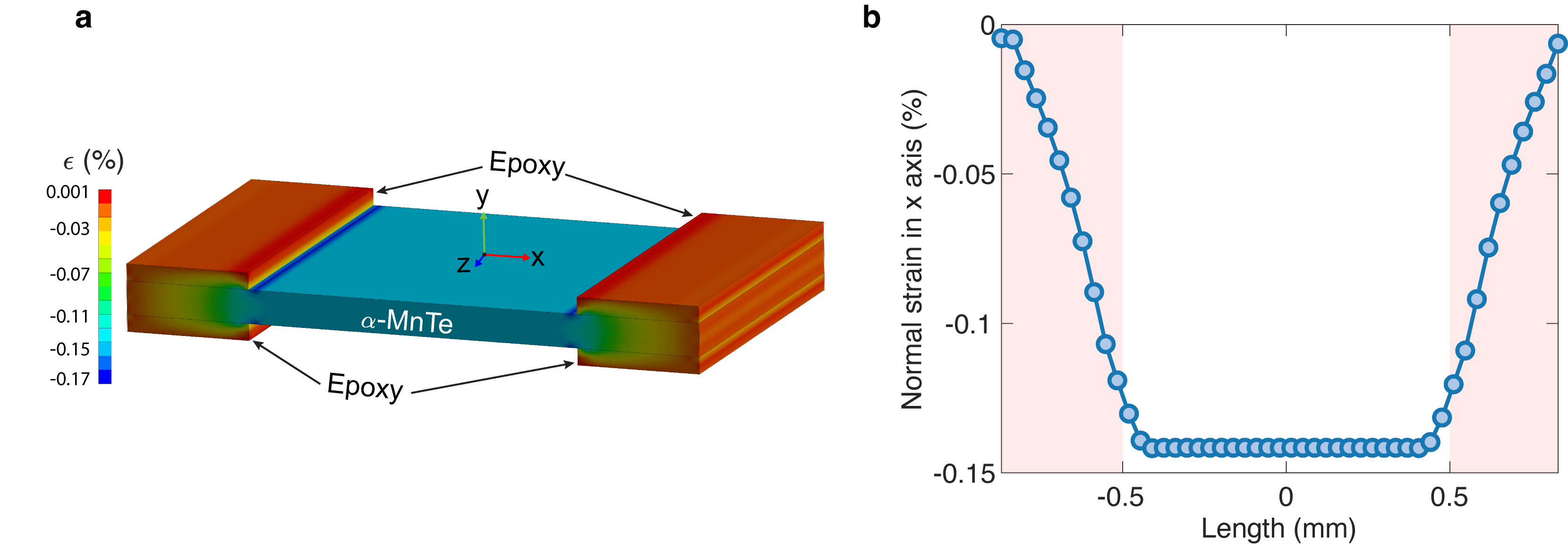}    
    \caption{\textbf{Finite element analysis of strain transmission.}  
    \textbf{a,} Simulated normal strain map in the epoxy and sample for a $-0.2\,\%$ nominal strain. The strain distribution within the sample, excluding the epoxy-covered regions, is nearly homogeneous. 
    \textbf{b,} Normal strain as function of position along the sample length ($x$ axis). The shaded area marks the epoxy-covered region. The central point of the sample reaches a strain of $-0.14\,\%$, corresponding to a strain transmission ratio of 0.7.
    \label{figS:FEA}}
\end{figure*}

\begin{figure*}
    \centering
    \includegraphics[width = 180mm]{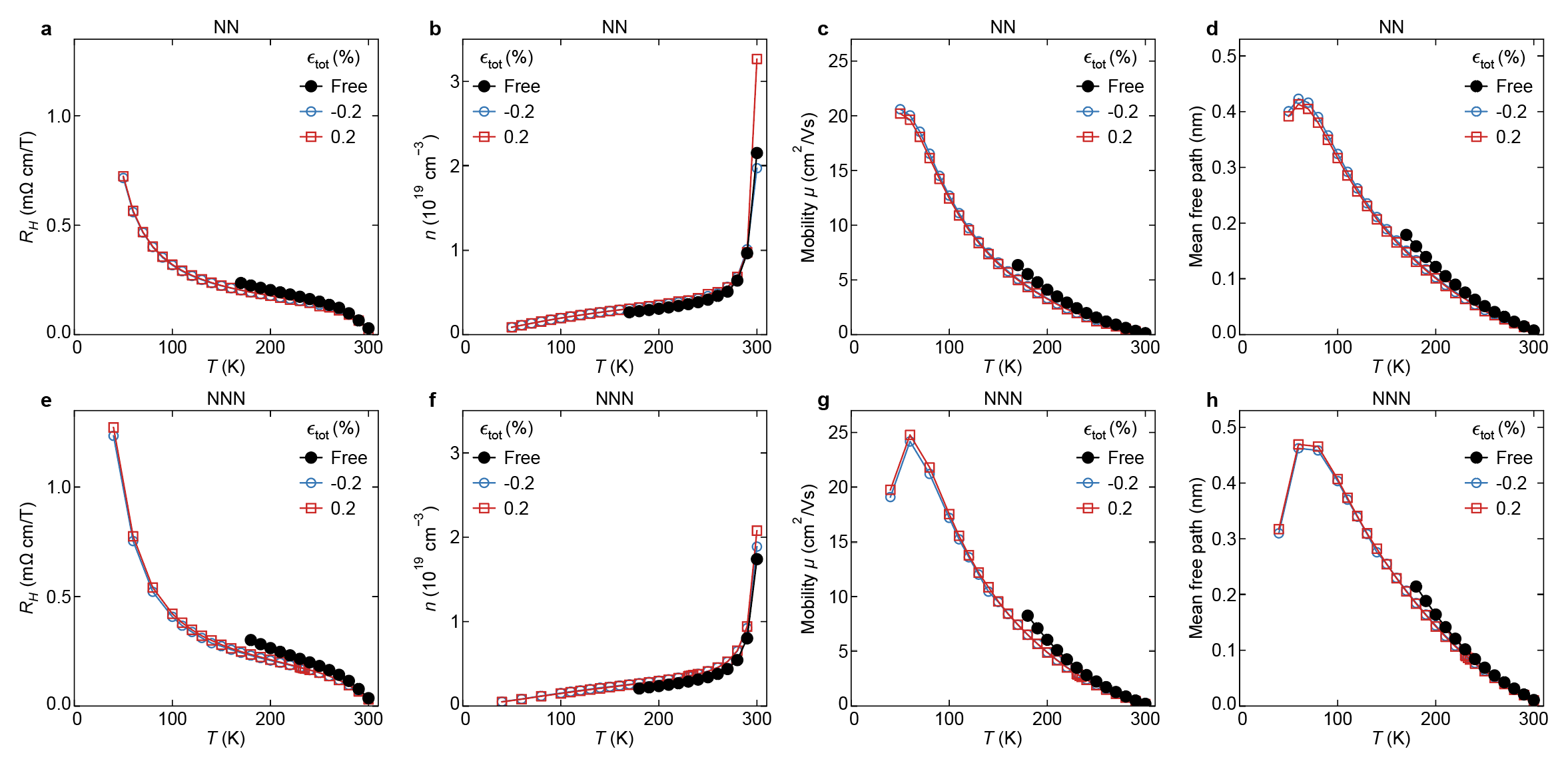}    
    \caption{\textbf{Temperature dependence of transport parameters in free-standing and strained $\alpha$-MnTe for two samples.}
    The top row (\textbf{a--d}) shows the sample measured in the NN-strain configuration in the main text, together with its corresponding free-standing dataset for comparison. The bottom row (\textbf{e--h}) shows the sample measured in the NNN-strain configuration in the main text, together with its corresponding free-standing dataset.
    \textbf{a,e,} Hall coefficient $R_H$.
    \textbf{b,f,} Carrier density $n$.
    \textbf{c,g,} Mobility $\mu$.
    \textbf{d,h,} Mean free path $\ell$.
    \label{figS:RH_n}}
\end{figure*}

\begin{figure*}
    \centering
    \includegraphics[width = 120mm]{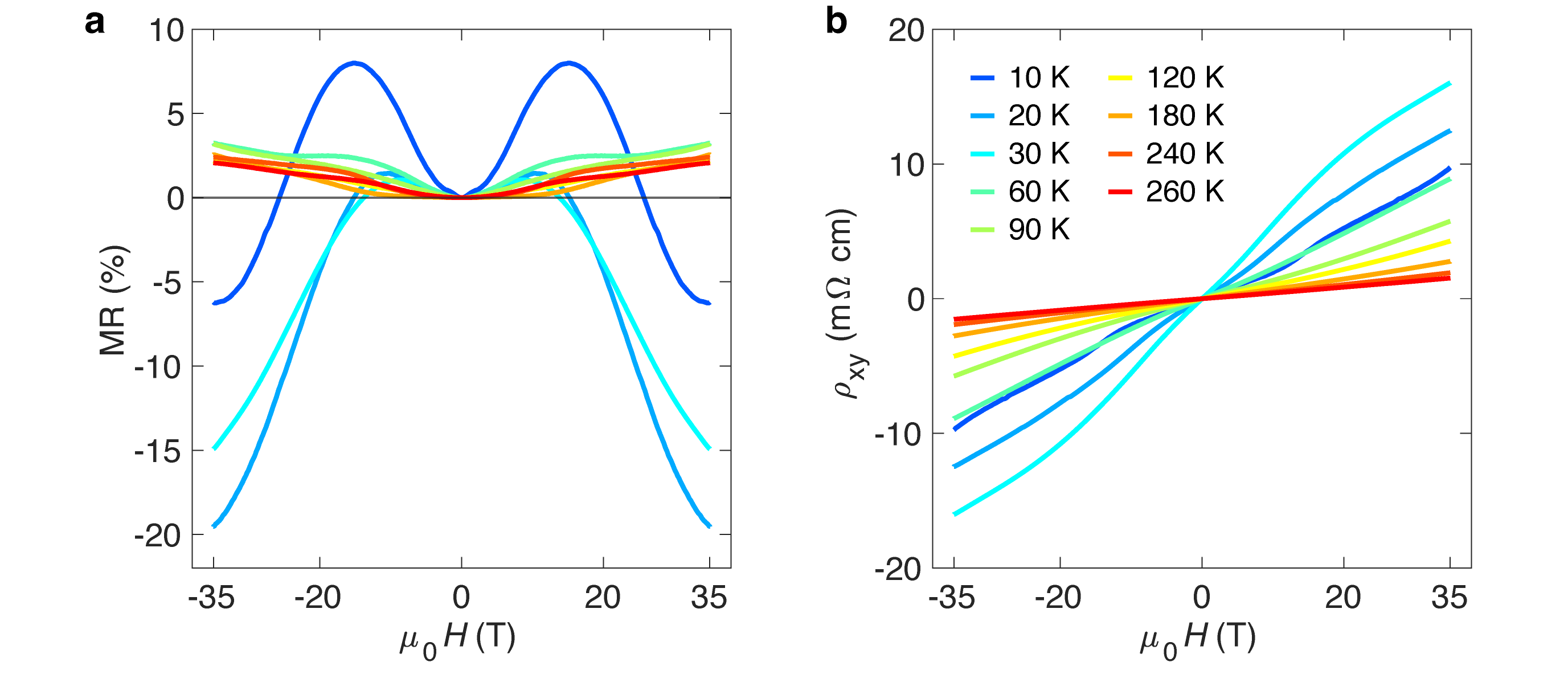}    
    \caption{\textbf{The magneto-resistance and Hall resistivity under high magnetic fields at various temperatures.}  
    \textbf{a,} The magneto-resistance remains positive in the low-field range at all measured temperatures. However, at low temperatures (below 90 K), it shows a strong drop beyond a certain field, indicating suppression of scattering under high magnetic fields.
    \textbf{b,} Hall resistivity keeps linearity up to 35 T above 90 K. However, a nonlinear two-band feature appears below 60 K.
    \label{figS:highfield}}
\end{figure*}



\end{document}